\documentclass[11pt,a4paper]{article}

\usepackage[lmargin=1.0in,rmargin=1.0in,bottom=1.0in,top=1.0in,twoside=False]{geometry}
\usepackage{hchang}
\usepackage{tcolorbox}
\usepackage{mathtools}
\usepackage{graphicx}
\usepackage{enumerate}
\usepackage{tikz}

\usepackage{amsmath}

\usetikzlibrary{fit,backgrounds,calc,positioning}
\usetikzlibrary{shapes}
\usetikzlibrary{decorations.pathmorphing}
\usetikzlibrary{decorations.pathreplacing, calligraphy}
\tikzset{snake it/.style={decorate, decoration=snake}}

\usepackage{caption}   
\usepackage{subcaption} \usepackage{thmtools}
\usepackage{float}
\usepackage[T1]{fontenc}
\usepackage{enumitem}

\usepackage{microtype}
\usepackage{comment}
\usepackage{cleveref}
\usepackage{etoolbox}
\makeatletter
\patchcmd{\@thm}{\@kernel@refstepcounter}{\refstepcounter}{}{}
\makeatother
\usepackage[normalem]{ulem}

\usepackage{stmaryrd}
\usepackage{marvosym}
\usepackage{booktabs}
\usepackage{multirow}
\usepackage{makecell}

\usepackage[ruled,linesnumbered,vlined]{algorithm2e}
\usepackage{dashrule}
\tcbuselibrary{breakable}
\RequirePackage[color=red,backgroundcolor=TODOcolor,textcolor=red,linecolor=lightgray,bordercolor=red,tickmarkheight=3pt,figwidth=0.5\linewidth,figheight=4cm,figcolor=TODOcolor]{todonotes}
\presetkeys{todonotes}{size=\tiny}{}

\newtheorem{lemma}{Lemma}[section]
\newtheorem{theorem}[lemma]{Theorem}
\newtheorem{corollary}[lemma]{Corollary}

\newtheorem{observation}[lemma]{Observation}
\newtheorem{claim}[lemma]{Claim}

\newtheorem{conjecture}{Conjecture}

\newcommand{\dist}{\mathsf{dist}}

\newcommand{\real}{\mathbb{R}}

\setuptodonotes{size=\normalsize}

\def\cA{\ensuremath{\mathcal{A}}}

\def\cD{\ensuremath{\mathcal{D}}}

\def\cF{\ensuremath{\mathcal{F}}}

\def\cK{\ensuremath{\mathcal{K}}}

\def\cS{\ensuremath{\mathcal{S}}}
\def\cT{\ensuremath{\mathcal{T}}}

\def\cX{\ensuremath{\mathcal{X}}}

\def\bdry{\partial}

\usepackage{multicol}

\begin{document}
\pagenumbering{gobble}
\hypersetup{pageanchor=false}

\title{Induced-Minor-Free Graphs Admit a Star-Based Separator of Size $\Tilde{O}(\sqrt{n})$}

\author{Arnold Filtser\thanks{\protect\parbox[t]{0.79\linewidth}{Bar-Ilan University, Israel.
    Email: \texttt{arnold.filtser@biu.ac.il}.
    Supported by the ISRAEL SCIENCE FOUNDATION (grant Nos.~1042/22 and~2021/26).
    Funded by the European Union (ERC, NonMetEmbed, 101301875).
    Views and opinions expressed are however those of the author(s) only
    and do not necessarily reflect those of the European Union or the
    European Research Council. Neither the European Union nor the
    granting authority can be held responsible for them.}\hfill
    \protect\parbox[t]{0.15\linewidth}{\vspace{0pt}\protect\includegraphics[width=\linewidth]{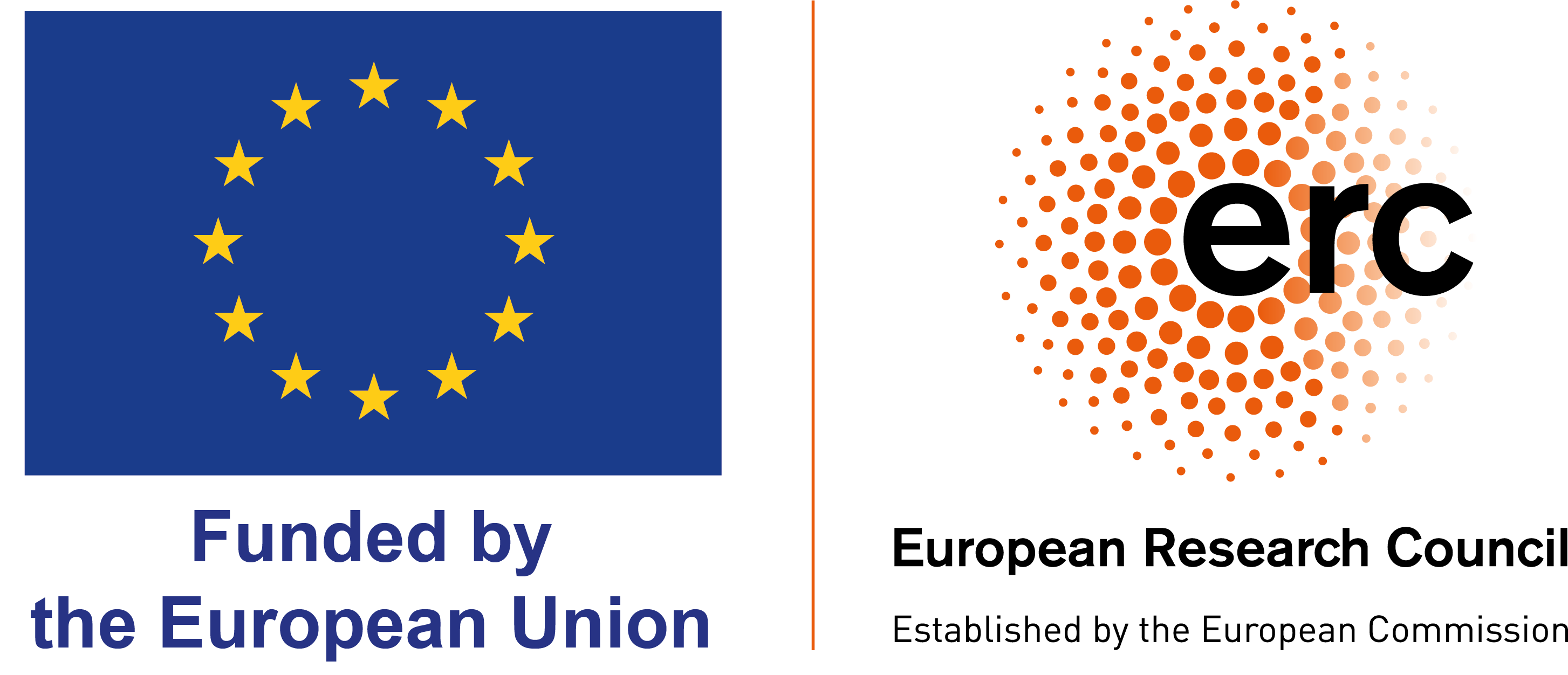}}}
  \and
  Nikolai Karol\thanks{Monash University, Australia.
    Email: \texttt{nikolai.karol@monash.edu}.}
  \and
  Hung Le\thanks{University of Massachusetts Amherst, USA.
    Email: \texttt{hungle@cs.umass.edu}.  Supported by the NSF CAREER award CCF-2237288 and the NSF grant CCF-2517033.}
  \and
  Florian Lehner\thanks{University of Auckland, New Zealand.
    Email: \texttt{florian.lehner@auckland.ac.nz}.}
}

\date{}

\makeatletter
\begingroup
\renewcommand{\@fnsymbol}[1]{\number#1}
\maketitle
\endgroup
\makeatother
\begin{abstract}

Given an $H$-induced-minor-free graph $G$ with $n$ vertices, we construct a balanced separator of $G$ consisting of $\tilde{O}(|V(H)|\sqrt{n})$ star subgraphs. Our result is an induced-minor-free analog of the famous separator theorem for $H$-minor-free graphs by Alon, Seymour, and Thomas~\cite{AST90}.
A key idea in our construction is a new decomposition of $H$-induced-minor-free graphs, analogous to the cop decomposition for $H$-minor-free graphs by Andreae~\cite {Andreae86}. We also discuss several algorithmic applications of our result.

\end{abstract}

\paragraph{AI Disclosure.} The key technical ideas and main results of this paper presented in \Cref{section:main}, specifically the decomposition as well as the star-based separator theorem, have been obtained entirely by the human authors. We used \EMPH{ChatGPT Astra} to assist with the algorithmic applications presented in \Cref{sec:applications}, specifically to collect several algorithmic applications. We then selected a few interesting problems and rewrote the section based on our understanding of the applications.  The human authors  wrote the entire paper, and ChatGPT was used to proofread. Human authors sketched the figures, and ChatGPT 5.6 converted them to TikZ.

\begingroup
\normalsize
\nolinenumbers
\setcounter{tocdepth}{2}
\setlength{\columnsep}{1.5em}
\setlength{\multicolsep}{6pt}
\vspace{30pt}
\begin{multicols}{2}[\noindent{\normalsize\bfseries\contentsname}]
\makeatletter
\@starttoc{toc}
\makeatother
\end{multicols}
\endgroup
\thispagestyle{empty}
\clearpage
\pagenumbering{arabic}
\hypersetup{pageanchor=true}

\section{Introduction} \label{sec:intro}

The Graph Minor Theory of Robertson and Seymour, developed over more than 20 years beginning in the early 1980s~\cite{RS83,RS10}, is the foundation of modern structural and algorithmic graph theory. Some of the most notable structural results with significant algorithmic consequences in this theory are the grid-minor theorem~\cite{RS86}, the Alon-Seymour-Thomas separator theorem~\cite{AST90}, and the Robertson-Seymour decomposition theorem~\cite{RS03}. 

We now overview these results. A graph $H$ is a \EMPH{minor} of a graph $G$ if $H$ can be obtained from $G$ by a sequence of vertex deletions, edge contractions, and edge deletions. A graph $G$ is \EMPH{$H$-minor-free} if $H$ is not a minor of $G$.
First, a seminal result of Robertson and Seymour~\cite{RS86} says that for every planar graph $H$, every $H$-minor-free graph has treewidth~\cite{RS84} at most some function $g(H)$. This result is often called the grid-minor theorem since it suffices to prove it when $H$ is a planar grid. We state the grid-minor theorem as \Cref{thm:sep-tw}, using an equivalent formulation phrased in terms of balanced separators. A \EMPH{balanced separator} of an $n$-vertex graph $G$ is a set $X \subseteq V(G)$ such that every connected component of $G - X$ has size at most $2n/3$. The \EMPH{size} of a balanced separator $X$ is the number of vertices in $X$, denoted $|X|$.

\begin{theorem}[Robertson-Seymour~\cite{RS86}]\label{thm:sep-tw} For every planar graph $H$ with $h$ vertices, every $H$-minor-free graph has a balanced separator of size $O_h(1)$, where $O_h$ hides a dependency on $h$.
\end{theorem}

\Cref{thm:sep-tw} has led to an explosion of treewidth-based graph algorithms over the last few decades.  The dependency on $h$ in the original grid minor theorem of Robertson and Seymour was extremely large, but a long line of work~\cite{RST94,KK20,LS15,CC16,C15,C16} has improved it to the state-of-the-art bound $O(h^{9}\polylog(h))$~\cite{CT21}. 

The second major result of the Graph Minor Theory is the separator theorem by Alon, Seymour, and Thomas~\cite{AST90}, which is a generalization of the famous Lipton-Tarjan planar separator theorem~\cite{LT79}.

\begin{theorem}[Alon-Seymour-Thomas~\cite{AST90}]\label{thm:AST-separator} For every graph $H$ with $h$ vertices, every $H$-minor-free graph $G$ has a balanced separator of size $O(h^{3/2}\sqrt{n})$.
\end{theorem}

The dependency on $h$ in \Cref{thm:AST-separator} is not optimal and has been studied by a number of papers~\cite{BLR10,KR10,SJ25,L26}. Very recently, an optimal bound of $O(h\sqrt{n})$ was obtained by~\cite{CLS26}.

Finally, the third result is the Robertson-Seymour decomposition theorem, which is perhaps the greatest achievement of the Graph Minor Theory. Stating it formally requires a lot of notation. In a nutshell, it says that every $H$-minor-free graph can be decomposed into pieces that can be nearly embedded in surfaces of bounded genus. This decomposition enabled numerous algorithmic results, including meta-algorithmic frameworks~\cite{G03,DHK05,DH08,DHK11}. 

\paragraph{Induced minors.}
A possible way to extend results of the classical Graph Minor Theory is to consider a relaxation of minors. One standard approach is the concept of induced minors. A graph $H$ is an \EMPH{induced minor} of a graph $G$ if $H$ can be obtained from $G$ by a sequence of vertex deletions and edge contractions. In comparison with the standard concept of `minor', this definition does not allow edge deletions in the sequence of operations. A graph $G$ is \EMPH{$H$-induced-minor-free} if $H$ is not an induced minor of $G$. We frequently say that a graph class $\mathcal{G}$ is \EMPH{minor-free} (or \EMPH{induced-minor-free}), if there exists a graph $H$ such that every graph in $\mathcal{G}$ is $H$-minor-free (or $H$-induced-minor-free, respectively).

Observe that if $H$ is an induced minor of $G$, then $H$ is a minor of $G$. So the class of $H$-minor-free graphs is a subclass of the class of $H$-induced-minor-free graphs. In contrast, induced-minor-free graph classes are much broader than minor-free graph classes. For example, while minor-free graph classes are sparse~\cite{K84,thomason1984extremal}, induced-minor-free graphs can be very dense: the complete graph $K_n$ is $P_2$-induced-minor-free for any $n$, where $P_2$ denotes the path of length $2$. In addition, induced-minor-free graphs capture many well-studied graph classes in computer science, such as \EMPH{string graphs} (intersection graphs of curves in the plane), that are not minor-free.

Inspired by the Graph Minor Theory of Robertson and Seymour, a recent line of research aims to establish an analogous theory for induced-minor-free graph classes. However, much less is known for induced-minor-free graph classes compared to minor-free classes. Perhaps the grandest goal is to establish a decomposition theorem for induced-minor-free graphs analogous to the Robertson-Seymour decomposition theorem mentioned above, but this seems to be far-fetched, if not impossible. Luckily, for the other two results, namely \Cref{thm:sep-tw} and \Cref{thm:AST-separator}, we at least have conjectural counterparts.

To formally state the conjectures, we need to introduce the notion of star-based separator proposed by de Berg, Jansen, and Lamme~\cite{BJL26}.  A graph $S$ is a \EMPH{star} if $S$ is isomorphic to $K_{1}$ or $K_{1,t}$ for some $t \geqslant 1$. Given a graph $G$ with $n$ vertices, let $\cF$ be a set of connected (not necessarily induced) subgraphs of $G$. We say that $\cF$ is a \EMPH{balanced star-based separator} of $G$ if (i) every connected component of $G\setminus (\cup_{F\in \cF}V(F))$  has size at most $2n/3$ and (ii) every $F\in \cF$ is a star. The \EMPH{size} of $\cF$, denoted \EMPH{$|\cF|$}, is the number of elements (stars) in~$\cF$.

\begin{conjecture}[Gartland-Lokshtanov Conjecture~\cite{Gartland23}]\label{conj:induced-planar} For every planar graph $H$ with $h$ vertices, every $H$-induced-minor-free graph $G$ has a balanced star-based\footnote{Star-based separators are also called \EMPH{dominated balanced separators} in some papers. Throughout this paper, we use the `star-based' terminology since it is more intuitive.} separator of size $O_h(1)$.
\end{conjecture}

This conjecture has received considerable attention and has been established for various special cases~\cite{CPPT24,DMS24,AGHK25,CCLMS25,CGKM25,CMPRR26}, but the full generality remains wide open. 

The second conjecture, posed by various authors~\cite{BonnetLPP26,BJL26} in different contexts, seeks a balanced separator analogous to \Cref{thm:AST-separator}.

\begin{conjecture}\label{conj:induced-minor} For every graph $H$ with $h$ vertices, every $H$-induced-minor-free $n$-vertex graph $G$ has a balanced star-based separator of size $O_h(\sqrt{n})$.
\end{conjecture}

\Cref{conj:induced-minor} is also wide open, even for string graphs~\cite{BJL26}. We note that the $\sqrt{n}$ bound in \Cref{conj:induced-minor} is asymptotically best possible, witnessed by the planar $\sqrt{n}\times\sqrt{n}$-grid. Korhonen and Lokshtanov~\cite{KH24} proved that for every graph $H$, every $H$-induced-minor-free $n$-vertex $m$-edge graph $G$ has a balanced (vertex) separator of size $O_H(\sqrt{m}\log m) = \Tilde{O}_H(\sqrt{m})$ (the $\Tilde{O}$ notation hides a polylog factor). We now show that this result implies that $G$ has a balanced star-based separator of size $\Tilde{O}_H(n^{2/3})$. Choose a parameter $\Delta > 0$. As long as the graph $G$ has a vertex $u$ of degree at least $\Delta$, add its neighborhood $N[u]$ (forming a star centered at $u$) to the separator, remove $N[u]$ from the graph, and recurse. Once every vertex has degree at most $\Delta$, the number of edges in the graph is $O(n\Delta)$, and by the Korhonen-Lokshtanov result there is a balanced (vertex) separator of size $\Tilde{O}_H(\sqrt{n\Delta})$.  The size of the final star-based separator is $n/(\Delta+1) + \Tilde{O}_H(\sqrt{n\Delta}) = \Tilde{O}_H(n^{2/3})$ by setting $\Delta = n^{1/3}$. A recent result~\cite{DaviesRSW26} removes the $\log n$ factor from the size of the Korhonen-Lokshtanov separator, and hence we can also remove the log factor from the $\Tilde{O}_H(n^{2/3})$ separator bound. Thus, $O_H(n^{2/3})$ remains the state-of-the-art bound; proving \Cref{conj:induced-minor} seems to require a completely different approach and new ideas.

\subsection{Our results}

Our main result is a resolution of \Cref{conj:induced-minor} up to a $\sqrt{\log n}$ factor.

\begin{restatable}{theorem}{separatorH}
\label{thm:separator-H}
For every graph $H$ with $h$ vertices, every $H$-induced-minor-free $n$-vertex graph $G$ has a balanced star-based separator of size $O(h\sqrt{n \log n})$.
\end{restatable}

Previous approaches for constructing balanced separators (not necessarily star-based) for string graphs and induced minor-free graphs rely heavily on flow-cut duality~\cite{M14,L17,KH24,DaviesRSW26}. Our proof of \Cref{thm:separator-H} relies on a different approach, inspired by the classical work of Plotkin, Rao, and Smith~\cite{PRS94} and the cop-and-robber game. 

In the \EMPH{cop-and-robber game} on a connected graph $G$, two players compete: a \EMPH{cop player} and a \EMPH{robber player}. The cop player controls $c$ cops, for some integer $c \geqslant 1$, and places each cop at a vertex of $G$; the cops can share their locations. Then the robber player places one robber at a vertex of $G$. In each round, the players take turns, and the cop player moves first. Each cop or robber could stay put or move to a new vertex along an edge. The cop player wins if a cop occupies the robber's vertex; in this case, we say that the cop \EMPH{captures} the robber. The \EMPH{cop number} of $G$, denoted by \EMPH{$c(G)$}, is the minimum number of cops such that the cop player has a winning strategy (regardless of the robber's strategy). 

Andreae~\cite{Andreae86} showed that every $H$-minor-free graph has a cop number bounded by $|E(H)|$; the proof gave a \EMPH{cop decomposition} of $H$-minor-free graphs. In a totally unrelated work, Plotkin, Rao, and Smith~\cite{PRS94} constructed a balanced separator for so-called shallow-minor-free graphs. The proof is very similar to the construction of the cop decomposition~\cite {Andreae86}, except at the step where the graph has large diameter. For $H$-minor-free graphs, the separator by Plotkin, Rao, and Smith~\cite{PRS94} has size $O(|V(H)|\sqrt{n\log n})$, which (unsurprisingly) is the same as the bound in our \Cref{thm:separator-H}.

Our proof of \Cref{thm:separator-H} adapts the strategy of Andreae~\cite{Andreae86} and Plotkin, Rao, and Smith~\cite{PRS94} to the induced-minor-free setting. Our proof constructs a new decomposition of induced-minor-free graphs, analogous to the cop decomposition for minor-free graphs by Andreae. \Cref{subsec:overview} presents an overview of our proof of \Cref{thm:separator-H} and a more detailed discussion on our decomposition of induced-minor-free graphs and its relation to the cop-and-robber game.

In \Cref{sec:applications} we describe several applications of our \Cref{thm:separator-H}. Consider an $H$-induced-minor-free connected $n$-vertex graph $G$, where a graph $H$ has $h$ vertices.
First, we bound the lazy cop number of $G$, where at most one cop moves in each turn, by $O(h\sqrt{n\log n})$ (\Cref{thm:lazy-cops}), generalizing  (up to second-order terms) the $O(\sqrt{n})$ bound of Bal et al.~\cite{Bal_2016} for graphs of bounded genus. We obtain distance oracles with additive stretch $2$, $\Tilde{O}(hn^{3/2})$ space, and $\Tilde{O}(h\sqrt{n})$ query time, improving on the oracle of de Berg, Jansen, and Lamme~\cite{BJL26}, which uses $O(n^{5/3}\log^{2/3}n)$ space and $O(n^{2/3}\log^{2/3}n)$ query time with the same additive stretch, and applies only to string graphs. Further, we obtain distance labels of $\Tilde{O}(h\sqrt{n})$ bits and spanning tree covers with $\Tilde{O}(h\sqrt{n})$ trees, both with additive stretch $2$; the latter yield path-reporting distance oracles with the same space and query bounds (\Cref{thm:additive-oracle,thm:tree-cover} and \Cref{cor:additive-labeling}). Finally, we construct embeddings with additive stretch $2$ into graphs of treewidth $O(h\sqrt{n\log n})$ (\Cref{thm:ai-lowtw-emulator}). Combined with existing algorithms for graphs of bounded treewidth~\cite{KLP19,KLP22,CCS25}, these embeddings yield $2^{\Tilde{O}(h\sqrt{n})}$-time additive approximation algorithms for $\rho$-dominating set, $\rho$-independent set, $k$-Center, and $k$-Median.

\section{Proof Overview}\label{subsec:overview}

In this section, we outline the proof of \Cref{thm:separator-H}. We denote the 1-subdivision of $K_t$ by \EMPH{$\dot K_t$}. It is well known that if $G$ excludes a graph $H$ with $h$ vertices as an induced minor, then $G$ excludes $\dot K_h$ as an induced minor. Thus, by proving the following theorem, we obtain \Cref{thm:separator-H} as a simple corollary.

\begin{restatable}{theorem}{separator}
\label{thm:separator}
If an $n$-vertex graph $G$ is $\dot K_t$-induced-minor-free,
then $G$ has a star-based balanced separator of size
$O(t\sqrt{n \log n})$.
\end{restatable}

We now give an overview of the proof of \Cref{thm:separator}. 
We use standard graph-theoretic terminology. For every vertex $u \in V(G)$, we denote by $N_G(u)$ the set of neighbors of $u$ in $G$ and $N_G[u] := N_G(u) \cup \{u\}$.
For each set $X \subseteq V(G)$, let $N_G(X) := \cup_{v \in X}N_G(v)$ and $N_G[X] := \cup_{v \in X}N_G[v]$. 
The subgraph of $G$ \EMPH{induced by} $X$, denoted \EMPH{$G[X]$}, has vertex set $X$ and its edge set is the set of $G$ edges with both endpoints in $X$.
Sometimes, for a subgraph $C$ of $G$, we will slightly abuse notation by using \EMPH{$G[C]$} to denote the induced subgraph $G[V(C)]$. 
For a subgraph $C$ of $G$, the \EMPH{boundary of $C$}, denoted \EMPH{$\bdry C$}, is the set of vertices in $C$ that have neighbors outside $C$. 
That is, $\bdry C := \{v\in V(C): N_G(v)\setminus V(C) \not=\varnothing\}$.

\begin{figure}[!htb]
\centering
\resizebox{0.9\textwidth}{!}{\begin{tikzpicture}[
    x=1cm,
    y=1cm,
    line cap=round,
    line join=round,
    ambientv/.style={
        circle,
        fill=black,
        inner sep=1.7pt
    },
    skelv/.style={
        circle,
        draw=red!80!black,
        fill=white,
        line width=1.2pt,
        minimum size=5.5mm,
        inner sep=0pt
    },
    centerv/.style={
        circle,
        draw=black,
        fill=white,
        line width=1.2pt,
        minimum size=5.5mm,
        inner sep=0pt
    },
    skedge/.style={
        red!85!black,
        line width=1.6pt
    },
    ambientedge/.style={
        black,
        line width=.7pt
    },
    modeledge/.style={
        blue!75!cyan,
        line width=2.1pt
    },
    absrouter/.style={
        rectangle,
        draw=red!80!black,
        fill=red!5,
        line width=1.1pt,
        minimum size=7.5mm
    },
    absconnector/.style={
        circle,
        draw=blue!75!cyan,
        fill=white,
        line width=1.1pt,
        minimum size=5.5mm,
        inner sep=0pt
    }
]

\node[absrouter] (A1) at (-7.20,1.60) {};
\node[absrouter] (A2) at (-4.10,1.60) {};
\node[absrouter] (A3) at (-5.65,0.35) {};
\node[absrouter] (A4) at (-5.65,-1.65) {};

\node[absconnector] (A12) at (-5.65,1.65) {};
\node[absconnector] (A13) at (-6.45,0.95) {};
\node[absconnector] (A23) at (-4.85,0.95) {};
\node[absconnector] (A14) at (-6.80,-0.05) {};
\node[absconnector] (A24) at (-4.50,-0.05) {};
\node[absconnector] (A34) at (-5.65,-0.65) {};

\draw[blue!75!cyan,line width=1.15pt] (A1)--(A12)--(A2);
\draw[blue!75!cyan,line width=1.15pt] (A1)--(A13)--(A3);
\draw[blue!75!cyan,line width=1.15pt] (A2)--(A23)--(A3);
\draw[blue!75!cyan,line width=1.15pt] (A1)--(A14)--(A4);
\draw[blue!75!cyan,line width=1.15pt] (A2)--(A24)--(A4);
\draw[blue!75!cyan,line width=1.15pt] (A3)--(A34)--(A4);

\node[font=\Large] at (-5.65,-2.45) {$\dot K_4$};

\draw[
    ->,
    >=stealth,
    orange!90!yellow,
    line width=2.7pt
]
(-3.35,0)
to[out=15,in=165]
(-2.10,0.15);

\begin{scope}[on background layer]
    \fill[red!10] (-0.90,1.00) rectangle (1.35,3.20);  \fill[red!10] (7.75,1.10) rectangle (9.45,3.35);   \fill[red!10] (4.40,-0.70) rectangle (5.85,0.95);  \fill[red!10] (4.25,-4.00) rectangle (6.35,-2.25); \end{scope}

\begin{scope}[on background layer]

\begin{scope}[shift={(4.65,2.55)},rotate=0]
    \fill[blue!10] (0,0) ellipse [x radius=.75cm,y radius=1.10cm];
\end{scope}

\begin{scope}[shift={(0.75,-0.50)},rotate=25]
    \fill[blue!10] (0,0) ellipse [x radius=.78cm,y radius=.47cm];
\end{scope}

\begin{scope}[shift={(8.10,-0.40)},rotate=-30]
    \fill[blue!10] (0,0) ellipse [x radius=.85cm,y radius=.47cm];
\end{scope}

\begin{scope}[shift={(2.45,0.80)},rotate=30]
    \fill[blue!10] (0,0) ellipse [x radius=.65cm,y radius=.42cm];
\end{scope}

\begin{scope}[shift={(6.95,1.10)},rotate=-25]
    \fill[blue!10] (0,0) ellipse [x radius=.62cm,y radius=.87cm];
\end{scope}

\begin{scope}[shift={(5.35,-1.42)},rotate=-10]
    \fill[blue!10] (0,0) ellipse [x radius=.58cm,y radius=.48cm];
\end{scope}

\end{scope}

\node[skelv] (r1a) at (0.00,2.70) {};
\node[skelv] (r1b) at (0.35,1.95) {};
\node[skelv] (r1c) at (-0.55,1.70) {};
\node[skelv] (r1d) at (0.35,1.25) {};

\node[ambientv] (r1x) at (-0.62,2.40) {};
\node[ambientv] (r1y) at (1.05,2.30) {};
\node[ambientv] (r1z) at (-0.35,1.22) {};

\draw[skedge] (r1a)--(r1c);
\draw[skedge] (r1a)--(r1b);
\draw[skedge] (r1b)--(r1d);

\draw[ambientedge] (r1x)--(r1a);
\draw[ambientedge] (r1a)--(r1y);
\draw[ambientedge] (r1z)--(r1b);
\draw[ambientedge] (r1z)--(r1d);

\node[text=red!80!black,font=\Large] at (1.55,3.55) {$\rho_1$};

\node[skelv] (r2a) at (8.00,2.05) {};
\node[skelv] (r2b) at (8.60,2.72) {};
\node[skelv] (r2c) at (8.30,1.28) {};
\node[skelv] (r2d) at (9.10,1.62) {};

\node[ambientv] (r2x) at (8.15,3.15) {};

\draw[skedge] (r2a)--(r2b);
\draw[skedge] (r2b)--(r2c);
\draw[skedge] (r2c)--(r2d);

\draw[ambientedge] (r2x)--(r2a);
\draw[ambientedge] (r2x)--(r2b);

\node[text=red!80!black,font=\Large] at (8.05,3.85) {$\rho_2$};

\node[skelv] (r4a) at (4.80,0.08) {};
\node[skelv] (r4b) at (5.45,0.58) {};
\node[skelv] (r4c) at (5.50,-0.20) {};

\node[ambientv] (r4x) at (4.85,-0.45) {};

\draw[skedge] (r4a)--(r4b);
\draw[skedge] (r4b)--(r4c);

\draw[ambientedge] (r4a)--(r4x);
\draw[ambientedge] (r4x)--(r4c);

\node[text=red!80!black,font=\Large] at (4.55,1.20) {$\rho_4$};

\node[skelv] (r3a) at (4.80,-2.62) {};
\node[skelv] (r3b) at (5.55,-3.00) {};
\node[skelv] (r3c) at (5.00,-3.68) {};

\node[ambientv] (r3x) at (4.45,-3.25) {};
\node[ambientv] (r3y) at (6.10,-3.20) {};

\draw[skedge] (r3a)--(r3b);
\draw[skedge] (r3b)--(r3c);

\draw[ambientedge] (r3x)--(r3a);
\draw[ambientedge] (r3x)--(r3b);
\draw[ambientedge] (r3y)--(r3b);

\node[text=red!80!black,font=\Large] at (6.82,-3.12) {$\rho_3$};

\node[centerv] (k1c) at (4.55,2.65) {};
\node[ambientv] (k1a) at (4.28,3.32) {};
\node[ambientv] (k1b) at (4.82,3.32) {};
\node[ambientv] (k1d) at (5.12,2.05) {};
\node[ambientv] (k1e) at (4.72,1.75) {};

\draw[ambientedge] (k1c)--(k1a);
\draw[ambientedge] (k1c)--(k1b);
\draw[ambientedge] (k1c)--(k1d);
\draw[ambientedge] (k1c)--(k1e);

\node[text=blue!80!cyan,font=\Large] at (5.02,3.95) {$\kappa_1$};

\node[centerv] (k2c) at (0.65,-0.60) {};
\node[ambientv] (k2a) at (0.80,0) {};

\draw[ambientedge] (k2c)--(k2a);

\node[text=blue!80!cyan,font=\Large] at (0.00,-1.30) {$\kappa_2$};

\node[centerv] (k3c) at (7.95,-0.50) {};
\node[ambientv] (k3a) at (8.50,-0.82) {};
\node[ambientv] (k3b) at (7.65,0.05) {};

\draw[ambientedge] (k3c)--(k3a);
\draw[ambientedge] (k3c)--(k3b);

\node[text=blue!80!cyan,font=\Large] at (8.55,-1.35) {$\kappa_3$};

\node[centerv] (k4c) at (2.45,0.80) {};
\node[ambientv] (k4a) at (2.05,0.45) {};

\draw[ambientedge] (k4c)--(k4a);

\node[text=blue!80!cyan,font=\Large] at (2.20,1.55) {$\kappa_4$};

\node[centerv] (k5c) at (7.02,1.10) {};
\node[ambientv] (k5a) at (6.70,1.57) {};
\node[ambientv] (k5b) at (6.75,0.38) {};

\draw[ambientedge] (k5c)--(k5a);
\draw[ambientedge] (k5c)--(k5b);

\node[text=blue!80!cyan,font=\Large] at (6.28,2.02) {$\kappa_5$};

\node[centerv] (k6c) at (5.35,-1.42) {};

\node[text=blue!80!cyan,font=\Large] at (6.05,-1.25) {$\kappa_6$};

\draw[modeledge] (r1b)--(k1c);
\draw[modeledge] (k1c)--(r2a);

\draw[modeledge] (r1d)--(k2c);
\draw[modeledge] (k2c)--(r3a);

\draw[modeledge] (r2c)--(k3c);
\draw[modeledge] (k3c)--(r3b);

\draw[modeledge] (r1b)--(k4c);
\draw[modeledge] (k4c)--(r4a);

\draw[modeledge] (r2a)--(k5c);
\draw[modeledge] (k5c)--(r4b);

\draw[modeledge] (r4c)--(k6c);
\draw[modeledge] (k6c)--(r3b);

\draw[ambientedge] (r1d)--(k4a);
\draw[ambientedge] (k4a)--(k2c);

\draw[ambientedge] (k1e)--(k5a);
\draw[ambientedge] (k5a)--(r2a);

\draw[ambientedge] (k2c)--(k6c);
\draw[ambientedge] (k6c)--(k5b);

\draw[ambientedge] (k5b)--(r2c);
\draw[ambientedge] (k6c)--(r3a);

\end{tikzpicture}
}
\caption{An induced model of $\dot K_4$ in a base graph $G$. The small picture on the left shows the abstract graph $\dot K_4$ being realized by the induced minor model $\cK$ on the right, with 10 supernodes. Each supernode is one of two types: routers or connectors.  The pale red regions are the four router supernodes $\rho_1,\rho_2,\rho_3,\rho_4$, and the red edges/white vertices inside them highlight their skeletons. The pale blue regions are the six connector supernodes $\kappa_1,\ldots,\kappa_6$, and the large white vertex inside each of them is its skeleton, which consists of a single vertex only. Thick blue edges indicate the required adjacencies from each connector center to the skeletons of its two associated routers.}
\label{fig:dotK4-supernode-model}
\end{figure}
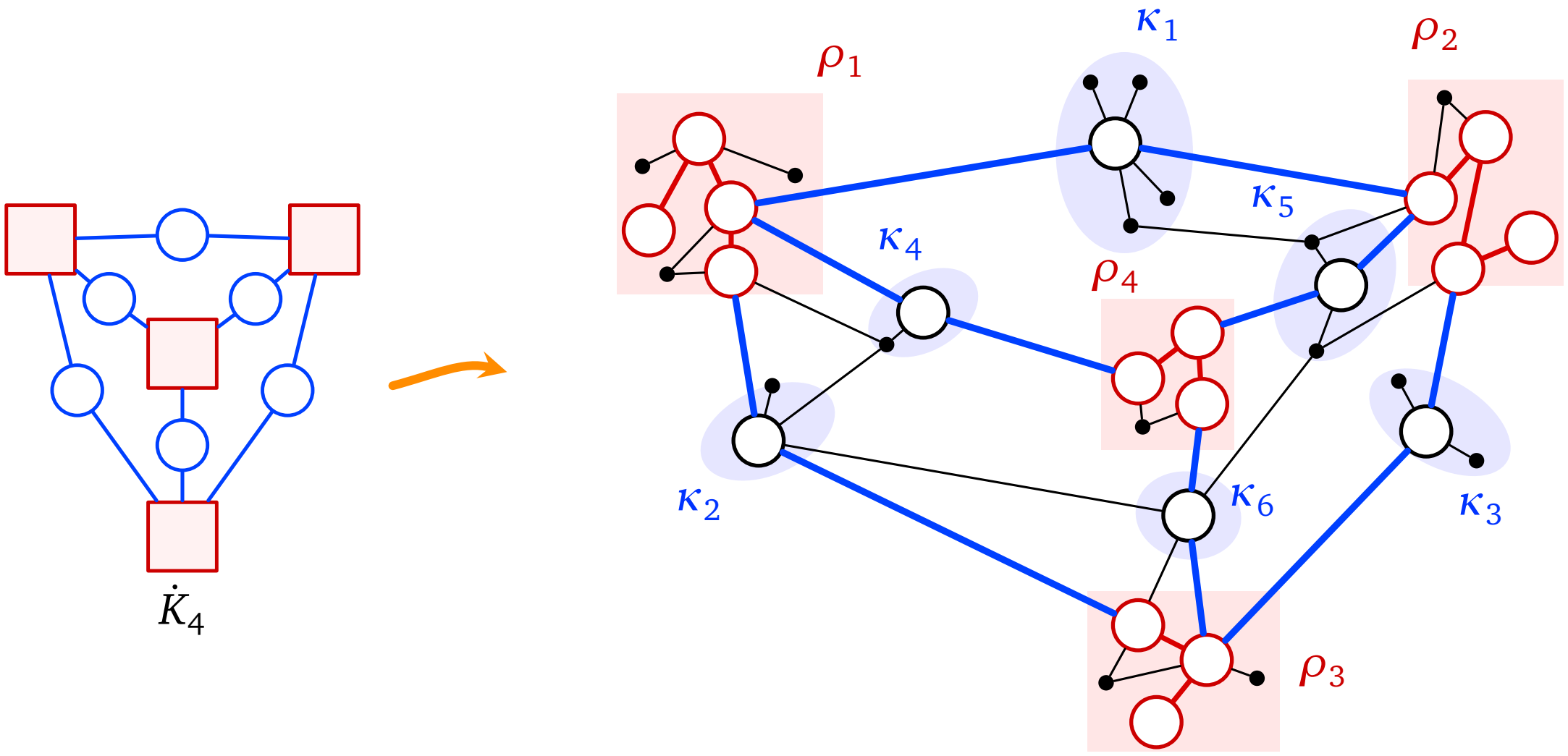

An important concept in our proof is an induced model of $\dot K_t$.  Let $T_{\eta}$ be a tree that is a subgraph of $G$. A \EMPH{supernode} $\eta$ with a \EMPH{skeleton} $T_{\eta}$ is a connected subgraph of $G[N_G[T_{\eta}]]$. 
That is, $\eta$ is a connected subgraph containing $T_{\eta}$ and some (if not all) neighbors of $T_{\eta}$ in $G$.  Given a collection of \underline{vertex-disjoint}\footnote{Strictly speaking, it suffices to have the vertex-disjointness between the skeletons of the supernodes in an induced minor model, but it is unclear whether relaxing vertex-disjointness would be more useful.} supernodes $\cK$ of $G = (V,E)$, we say that $\cK$ is an \EMPH{induced model} of $\dot K_t$ (see \Cref{fig:dotK4-supernode-model}) if:

\begin{itemize}
    \item \textbf{[Vertex correspondence]:} there is a one-to-one correspondence between each supernode of $\cK$ and a vertex of  $\dot K_t$. In particular, $\cK$ has exactly $t + {t\choose 2} = \frac{t(t+1)}{2}$ supernodes.
    \item  \textbf{[Edge correspondence]:} for every edge $ab\in \dot K_t$, its two corresponding supernodes, say $\eta_a$ and $\eta_b$, have an edge   $(x_a, x_b)$ in $G$ between their skeletons: $x_a\in T_{\eta_a}$ and $x_b\in T_{\eta_b}$.
    \item \textbf{[Non-edge correspondence]:}  If there is no edge between two vertices $a$ and $b$ in $\dot K_t$, then there is no edge between \underline{the skeletons} $T_{\eta_a}$ and   $T_{\eta_b}$ of their corresponding supernodes $\eta_a$ and~$\eta_b$.
\end{itemize}

Note that the non-edge correspondence property only requires no edge between the skeletons; there could still be an edge between the two supernodes $\eta_a$ and $\eta_b$, for example, between a vertex of $T_{\eta_a}$  and a vertex in $\eta_b\setminus T_{\eta_b}$. 

Observe that the existence of an induced model $\cK$ of $\dot K_t$ implies that $G$ contains  $\dot K_t$ as an induced minor: we can obtain $\dot K_t$ from $G$ by a sequence consisting of deletions of vertices of $G$ not in the skeleton of any supernode in $\cK$ and then contractions of the edges of the skeletons (of the  supernodes in $\cK$), each skeleton into a single vertex. So no edge deletions are needed.  
For a collection $\cK$ of supernodes, we denote by $V(\cK) := \bigcup_{\eta\in\cK}V(\eta)$ the set of all vertices in these supernodes.

\paragraph{The algorithm.} Our algorithm that constructs the balanced star-based separator can be described roughly as follows. It is an iterative algorithm where, in each step, we maintain a (partial) induced model $\cK$, a (partial) separator $I\subseteq V$, and an induced subgraph $C$ of $G$ such that all neighbors of $\bdry C$ outside of $C$ lie in $V(\cK)\cup I$, and $|V(C)|\geq 2n/3$. Intuitively, $C$ is the largest connected component of $G$ if we remove the ``current separator'' from $G$. (The precise definition of ``current separator'' is not very important for now.) Note that $I$ contains only graph vertices (not subgraphs). (At the outset of the algorithm, $C = G$, $I =\varnothing$ and $\cK = \varnothing$.) In every step, we shrink $C$ by considering two (big) cases:
\begin{itemize}
    \item \textbf{[Case 1]}: $C$ has radius at least $\Tilde{\Omega}(\sqrt{n})$. Then we find a subset of vertices $X\subseteq V(C)$ such that we can remove at least $\Omega(\sqrt{n}|X|)$ vertices out of $C$ in the next step. Then we add $X$ to $I$, and we set $C$ to be the largest connected component of $C\setminus X$. Since every time we add a vertex to $I$, we remove at least $\Tilde{\Omega}(\sqrt{n})$ vertices out of $C$, $|I| = \Tilde{O}(\sqrt{n})$ through the entire process.
    
    \item\textbf{[Case 2]}: $C$ has radius $\Tilde{O}(\sqrt{n})$. Then we add to $\cK$ a new supernode $\eta$, whose skeleton is a shortest path tree towards (some of) the previously created supernodes.
    The number of leaves in this skeleton is bounded by $t$ (as we can use these skeletons to create a minor model).
    Accordingly, we can break $\eta$ into $\Tilde{O}(t\cdot \sqrt{n})$ stars (using the radius bound).
    Then we set $C$ to be the largest connected component of $C\setminus \eta$.
\end{itemize}

The most delicate step, where our major technical innovation lies, is in Case 2: how to add a supernode to $\cK$ while respecting the edge correspondence and non-edge correspondence properties. For this purpose, we split the supernodes of $\cK$ into two types: \EMPH{routers} and \EMPH{connectors}. (Case 2 in our actual implementation is also different; specifically, we split it into two cases.) Routers correspond to vertices of the unsubdivided $K_t$ and connectors correspond to the subdivision (degree-2) vertices of $\dot K_t$. The role of connectors is to connect routers. See the illustration in \Cref{fig:dotK4-supernode-model}.  Managing routers and connectors is delicate: sometimes we move vertices from one supernode to another, and at other times we keep so-called \EMPH{ghost connectors}, formed to connect two routers, but one is removed as the algorithm proceeds.  To keep things relatively simple and high-level, we unfortunately have to sweep these details under the rug in this overview. We believe the concepts of routers and connectors, and the process to manage them, are of independent interest, since they give us a way to certify whether $G$ is induced-minor-free. In general, algorithms for induced-minor-free graphs need such a certificate.

As long as both cases are implemented correctly, we stop the algorithm whenever $|V(C)|< 2n/3$. The separator is \EMPH{$\cF =\textsc{BreakStar}(\cK) \cup I$}, where $\EMPH{\textsc{BreakStar}}(\cK)$ breaks supernode $\eta \in \cK$ into stars and collects all the resulting stars. Since the skeleton of each supernode has radius at most $\sqrt{n}$ and $t$ leaves, each supernode can be broken into $\Tilde{O}(t\sqrt{n})$ stars. Note that $I$ only has $\Tilde{O}(\sqrt{n})$ vertices, as noted in Case 1. Thus, the total number of stars in the separator is $\Tilde{O}(|\cK|t\sqrt{n}) = \Tilde{O}(t^3\sqrt{n})$. 

In our algorithm, to get $\Tilde{O}(t\sqrt{n})$ stars, we first tweak the radius threshold to be $\Tilde{O}(\sqrt{n}/t)$ instead of $\Tilde{O}(\sqrt{n})$; this tweak gives $\Tilde{O}(|\cK|t\frac{\sqrt{n}}{t}) = \Tilde{O}(t^2\sqrt{n})$. To save the last factor of $t$, we exploit the asymmetry between connectors and routers. Specifically, for each connector, we can reduce its skeleton to a single vertex (and hence each connector becomes a star). Then, the number of stars is:
\begin{equation*}
    \Tilde{O}\left(|\text{routers}|t(\frac{\sqrt{n}}{t}) + |\text{connectors}|\right)  = \Tilde{O}(t\sqrt{n} + t^2) = \Tilde{O}(t\sqrt{n})~,
\end{equation*}
since we only have $O(t)$ routers and $O(t^2)$ connectors. Here we use the fact that $t\leq \sqrt{n}$, since otherwise $t\sqrt{n} \geq n$, and $V(G)$ is a trivial separator of that size. We note that to get the final nearly optimal bound $\Tilde{O}(t\sqrt{n})$, we do not have to change the algorithm in any significant way.

As mentioned in \Cref{sec:intro}, our proof of \Cref{thm:separator-H} adapts the strategy of Andreae~\cite{Andreae86}, who showed that every connected $H$-minor-free graph has cop number bounded in terms of $H$. His proof gave a \EMPH{cop decomposition} of $H$-minor-free graphs. It is thus natural to ask how our decomposition relates to the cop-and-robber game.
Consider the \EMPH{distance-$1$} variant of the cop-and-robber game, in which a cop at a vertex $u$ \EMPH{captures} a robber at a vertex $v$ if $v=u$ or $v$ is a neighbor of $u$. The \EMPH{distance-$1$ cop number} of a graph $G$, denoted \EMPH{$c_1(G)$}, is the minimum number of cops such that the cop player has a winning strategy.
Recall that $c(G)$ denotes the classical cop number.
Denote by $G^{(3)}$ the graph obtained from $G$ by replacing every edge with a path of three edges.
It is known~\cite{BCP10,JKT10} that for every connected graph $G$,
$c(G)\le c_1(G^{(3)})\le c(G)+1$.
In particular, for every subdivision-closed family $\cF$ of connected graphs (like $H$-minor-free graphs),
$\sup_{G\in\cF}c(G)=\sup_{G\in\cF}c_1(G)$.\footnote{Indeed,
$\sup_{G\in\cF}c_1(G)\le\sup_{G\in\cF}c(G)
\le\sup_{G\in\cF}c_1(G^{(3)})
\le\sup_{G\in\cF}c_1(G)$,
where the last inequality follows since $G^{(3)}\in\cF$.}
Families of $H$-induced-minor-free graphs are not closed under subdivision, motivating the following conjecture.

\begin{conjecture} \label{conj:copsandrobbers} 
Let $\cF$ be the family of connected $H$-induced-minor-free graphs. Then $\sup_{G\in\cF}c_1(G)=O_{|H|}(1)$.
\end{conjecture}

It is natural to consider whether our decomposition resolves \Cref{conj:copsandrobbers}. However, \Cref{conj:copsandrobbers} remains open.
The decomposition (produced by our algorithm) does not solve \Cref{conj:copsandrobbers} due to a very specific technical detail we omitted here.\footnote{Andreae's~\cite{Andreae86} strategy is based on the fact that a single cop is sufficient to guard a single shortest path (regardless of its length). 
In our construction, initially, similarly to \cite{Andreae86}, the skeletons consist of shortest paths.
Unfortunately, during the delicate maintenance, this shortest-path property breaks, so the skeletons cannot be easily guarded.} For readers who are familiar with the cop-and-robber game, the maintenance of the big connected component $C$ in the algorithm is analogous to the \EMPH{robber area} in constructing a winning strategy for the cop player.

\section{Star-Based Separators for Induced-Minor-Free Graphs}
\label{section:main}

In this section, we prove \Cref{thm:separator}, restated below. As discussed in \Cref{subsec:overview}, \Cref{thm:separator} immediately implies \Cref{thm:separator-H}. We adopt the terminology outlined in \Cref{subsec:overview}. 

\separator*

\paragraph{Setting things up.~}  Recall that $\eta$ denotes a supernode, and $T_{\eta}$ denotes its skeleton;  $\eta$ is a connected subgraph containing $T_{\eta}$ and some (if not all) neighbors of $T_{\eta}$ in $G$.  
Let $\eta^+ := G[N_G[T_{\eta}]]$, so $\eta^+$ includes all vertices in $T_{\eta}$ and their neighbors. 
We will abuse notation and treat $\eta,\eta^+$ and $T_\eta$ both as subgraphs, and subsets of vertices.
Borrowing the terminology from the cop-and-robber game, we say that the 1-neighborhood of $T_{\eta}$ (which is exactly $V(\eta^+)$) is \EMPH{guarded} by $T_{\eta}$. 
For any set of supernodes $\cA$, denote $\cA^+ := \cup_{\eta \in \cA}\eta^+$, which is the set of all vertices guarded by the skeletons of supernodes in~$\cA$.

We now outline key definitions in our construction of supernodes given below. We will give each supernode a label: \EMPH{router} or \EMPH{connector}. 
The routers correspond to the original clique vertices in $\dot{K}_t$, while the connectors correspond to the subdivision vertices.
Each connector could be one of three types: \EMPH{open connector}, \EMPH{closed connector}, or \EMPH{ghost connector}. 
Each open connector is associated with exactly one router.  
Each closed connector is associated with exactly two routers, and the role of a closed connector is to connect these two different routers; the precise definition of ``connect'' will be made clear below. 
Open connectors arise in our attempt to connect one router to a future router, which will be formed in a future step. 
A ghost connector was previously a closed connector, but then it ``loses'' one of its associated routers during the process; some ghost connectors will later be turned into open connectors.  
Every connector $\kappa$ has the property that \textcolor{magenta}{its skeleton $T_{\kappa}$ is a single vertex}, called the \EMPH{center}.  

Let $\mathcal{K}$ be a set of supernodes. As illustrated in \Cref{fig:dotK4-supernode-model}, $\mathcal{K}$ is an \EMPH{induced model} of $\dot K_{t}$ if:
\begin{enumerate}
    \item $\mathcal{K}$ has exactly $t$ routers and $t \choose 2$ closed connectors. Each closed connector is associated with exactly two routers. For every two distinct routers $\rho_i$ and $\rho_j$, there is exactly one (closed) connector associated with $\rho_i$ and $\rho_j$.
    \item For every closed connector $\kappa$ associated with two routers $\rho_i$ and $\rho_j$, the center of $\kappa$ is adjacent to the skeletons of both $\rho_i$ and $\rho_j$.
    \item Except for the adjacency described in item 2 above, for every two supernodes (which are two connectors, two routers, or one connector and one router) in $\cK$, their skeletons are non-adjacent. 
\end{enumerate}

In an induced model $\cK$ of $\dot K_t$, each closed connector corresponds to a subdivision vertex of $\dot K_t$; items 2 and 3 above imply that each connector connects exactly two routers. Observe that an induced model of $\dot K_{t}$ in $G$ is a certificate that $G$ contains $\dot K_{t}$ as an induced minor.

\paragraph{Invariants of the construction.} We construct a star-based balanced separator iteratively. We will parameterize the size of the separator by $\ell \geq 1$, and construct a star-based balanced separator of size:

\begin{equation}\label{eq:sep-size}
    O(t^2\ell \log(n) + n/\ell)
\end{equation}
Then by choosing $\ell = \sqrt{n}/(t\sqrt{\log n})$, we obtain a separator that satisfies the conditions of \Cref{thm:separator}.

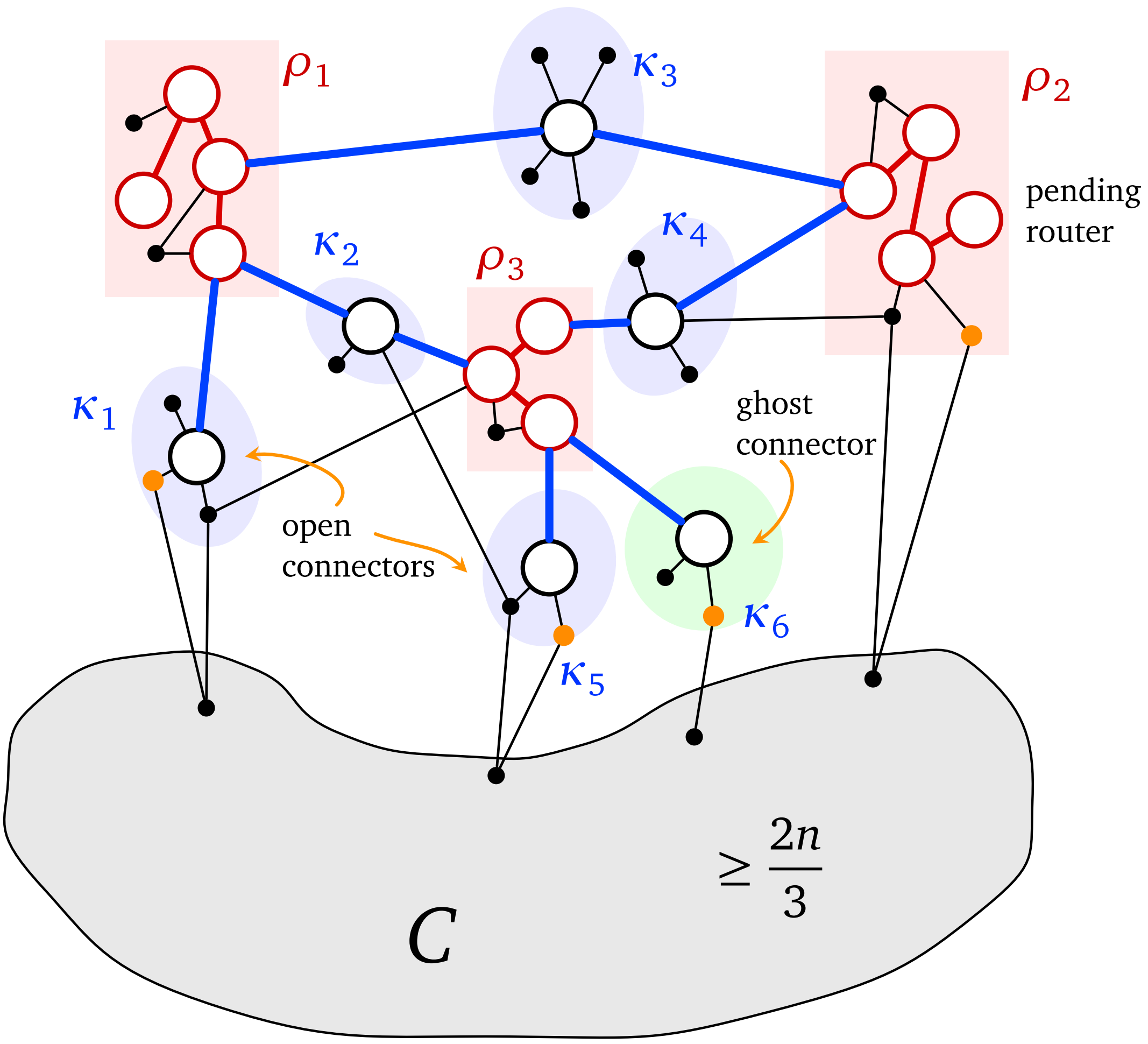
\begin{figure}[!htb]
\centering
\resizebox{0.9\textwidth}{!}{\begin{tikzpicture}[
    x=1cm,
    y=1cm,
    line cap=round,
    line join=round,
    ambientv/.style={
        circle,
        fill=black,
        inner sep=1.8pt
    },
    orangev/.style={
        circle,
        fill=orange!90!yellow,
        inner sep=2.2pt
    },
    skelv/.style={
        circle,
        draw=red!80!black,
        fill=white,
        line width=1.3pt,
        minimum size=5.5mm,
        inner sep=0pt
    },
    centerv/.style={
        circle,
        draw=black,
        fill=white,
        line width=1.3pt,
        minimum size=5.5mm,
        inner sep=0pt
    },
    skedge/.style={
        red!85!black,
        line width=1.7pt
    },
    ambientedge/.style={
        black,
        line width=.75pt
    },
    modeledge/.style={
        blue!75!cyan,
        line width=2.4pt
    },
    routerlabel/.style={
        text=red!80!black,
        font=\Large
    },
    connectorlabel/.style={
        text=blue!80!cyan,
        font=\Large
    }
]

\begin{scope}[on background layer]
\filldraw[
    fill=gray!18,
    draw=black,
    line width=.7pt
]
plot[smooth cycle, tension=0.9] coordinates {
    (-1.65,-3.10)
    (-1.25,-2.25)
    (-0.20,-1.85)
    (0.90,-2.00)
    (1.75,-2.65)
    (3.05,-2.90)
    (4.30,-2.78)
    (5.65,-2.15)
    (7.30,-1.85)
    (8.45,-2.05)
    (8.95,-3.35)
    (8.35,-4.55)
    (6.20,-5.55)
    (3.55,-5.80)
    (0.55,-5.50)
    (-1.30,-4.25)
};
\end{scope}

\node[font=\Huge] at (2.75,-4.75) {$C$};

\node[
    font=\Large,
    align=center
] at (6.25,-4.05)
{$\displaystyle \geq \frac{2n}{3}$};

\begin{scope}[on background layer]

\fill[red!9]
(-0.65,1.85) rectangle (1.15,4.5);

\fill[red!9]
(6.80,1.25) rectangle (8.70,4.40);

\fill[red!9]
(3.10,0.05) rectangle (4.40,1.95);

\end{scope}

\begin{scope}[on background layer]

\begin{scope}[shift={(0.30,0.20)},rotate=15]
    \fill[blue!9]
        (0,0) ellipse [x radius=.65cm,y radius=.95cm];
\end{scope}

\begin{scope}[shift={(2.05,1.50)},rotate=-35]
    \fill[blue!9]
        (0,0) ellipse [x radius=.68cm,y radius=.48cm];
\end{scope}

\begin{scope}[shift={(4.15,3.75)},rotate=0]
    \fill[blue!9]
        (0,0) ellipse [x radius=.78cm,y radius=1.10cm];
\end{scope}

\begin{scope}[shift={(5.20,1.75)},rotate=-20]
    \fill[blue!9]
        (0,0) ellipse [x radius=.65cm,y radius=.95cm];
\end{scope}

\begin{scope}[shift={(3.95,-0.95)},rotate=-15]
    \fill[blue!9]
        (0,0) ellipse [x radius=.68cm,y radius=.82cm];
\end{scope}

\begin{scope}[shift={(5.55,-0.75)},rotate=-5]
    \fill[green!12]
        (0,0) ellipse [x radius=.82cm,y radius=.85cm];
\end{scope}

\end{scope}

\node[skelv] (r1a) at (0.25,3.95) {};
\node[skelv] (r1b) at (0.55,3.20) {};
\node[skelv] (r1c) at (-0.25,2.85) {};
\node[skelv] (r1d) at (0.52,2.30) {};

\node[ambientv] (r1x) at (-0.35,3.65) {};
\node[ambientv] (r1y) at (-0.12,2.30) {};

\draw[skedge] (r1a)--(r1c);
\draw[skedge] (r1a)--(r1b);
\draw[skedge] (r1b)--(r1d);

\draw[ambientedge] (r1x)--(r1a);
\draw[ambientedge] (r1y)--(r1b);
\draw[ambientedge] (r1y)--(r1d);

\node[routerlabel] at (1.45,4.20) {$\rho_1$};

\node[skelv] (r2a) at (7.25,2.95) {};
\node[skelv] (r2b) at (7.90,3.55) {};
\node[skelv] (r2c) at (7.65,2.25) {};
\node[skelv] (r2d) at (8.35,2.65) {};

\node[ambientv] (r2x) at (7.35,3.95) {};
\node[ambientv] (r2y) at (7.50,1.65) {};
\node[orangev]  (r2o) at (8.32,1.45) {};

\draw[skedge] (r2a)--(r2b);
\draw[skedge] (r2b)--(r2c);
\draw[skedge] (r2c)--(r2d);

\draw[ambientedge] (r2x)--(r2a);
\draw[ambientedge] (r2x)--(r2b);

\draw[ambientedge] (r2c)--(r2y);
\draw[ambientedge] (r2c)--(r2o);

\node[routerlabel] at (9.10,4.05) {$\rho_2$};

\node[
    font=\small,
    align=left,
    anchor=west
] at (8.75,2.75)
{pending\\router};

\node[skelv] (r3a) at (3.90,1.55) {};
\node[skelv] (r3b) at (3.35,1.05) {};
\node[skelv] (r3c) at (3.95,0.55) {};

\node[ambientv] (r3x) at (3.40,0.45) {};

\draw[skedge] (r3a)--(r3b);
\draw[skedge] (r3b)--(r3c);

\draw[ambientedge] (r3x)--(r3b);
\draw[ambientedge] (r3x)--(r3c);

\node[routerlabel] at (3.45,2.20) {$\rho_3$};

\node[centerv] (k1c) at (0.30,0.20) {};
\node[ambientv] (k1a) at (0.05,0.75) {};
\node[ambientv] (k1b) at (0.42,-0.40) {};
\node[orangev]  (k1o) at (-0.15,-0.05) {};

\draw[ambientedge] (k1c)--(k1a);
\draw[ambientedge] (k1c)--(k1b);
\draw[ambientedge] (k1c)--(k1o);

\node[connectorlabel] at (-0.75,0.65) {$\kappa_1$};

\node[centerv] (k2c) at (2.10,1.55) {};
\node[ambientv] (k2a) at (1.75,1.15) {};

\draw[ambientedge] (k2c)--(k2a);

\node[connectorlabel] at (1.75,2.35) {$\kappa_2$};

\node[centerv] (k3c) at (4.15,3.60) {};
\node[ambientv] (k3a) at (3.85,4.35) {};
\node[ambientv] (k3b) at (4.55,4.35) {};
\node[ambientv] (k3d) at (3.75,3.10) {};
\node[ambientv] (k3e) at (4.28,2.75) {};

\draw[ambientedge] (k3c)--(k3a);
\draw[ambientedge] (k3c)--(k3b);
\draw[ambientedge] (k3c)--(k3d);
\draw[ambientedge] (k3c)--(k3e);

\node[connectorlabel] at (5.05,4.20) {$\kappa_3$};

\node[centerv] (k4c) at (5.05,1.60) {};
\node[ambientv] (k4a) at (4.85,2.25) {};
\node[ambientv] (k4b) at (5.40,1.05) {};

\draw[ambientedge] (k4c)--(k4a);
\draw[ambientedge] (k4c)--(k4b);

\node[connectorlabel] at (5.35,2.55) {$\kappa_4$};

\node[centerv] (k5c) at (3.95,-0.95) {};
\node[ambientv] (k5a) at (3.55,-1.35) {};
\node[orangev]  (k5o) at (4.10,-1.65) {};

\draw[ambientedge] (k5c)--(k5a);
\draw[ambientedge] (k5c)--(k5o);

\node[connectorlabel] at (4.30,-2.10) {$\kappa_5$};

\node[centerv] (k6c) at (5.55,-0.65) {};
\node[ambientv] (k6a) at (5.15,-1.05) {};
\node[orangev]  (k6o) at (5.65,-1.45) {};

\draw[ambientedge] (k6c)--(k6a);
\draw[ambientedge] (k6c)--(k6o);

\node[connectorlabel] at (6.20,-1.50) {$\kappa_6$};

\draw[modeledge] (r1b)--(k3c);
\draw[modeledge] (k3c)--(r2a);

\draw[modeledge] (r1d)--(k2c);
\draw[modeledge] (r1d)--(k1c);

\draw[modeledge] (k2c)--(r3b);

\draw[modeledge] (r3a)--(k4c);
\draw[modeledge] (k4c)--(r2a);

\draw[modeledge] (r3c)--(k5c);
\draw[modeledge] (r3c)--(k6c);

\node[ambientv] (C1) at (0.40,-2.40) {};
\draw[ambientedge] (k1o)--(C1);
\draw[ambientedge] (k1b)--(C1);

\node[ambientv] (C2) at (3.40,-3.10) {};
\draw[ambientedge] (k5a)--(C2);
\draw[ambientedge] (k5o)--(C2);

\node[ambientv] (C3) at (5.45,-2.70) {};
\draw[ambientedge] (k6o)--(C3);

\node[ambientv] (C4) at (7.30,-2.10) {};
\draw[ambientedge] (r2y)--(C4);
\draw[ambientedge] (r2o)--(C4);

\draw[ambientedge] (k1b)--(r3b);
\draw[ambientedge] (k2c)--(k5a);
\draw[ambientedge] (k4c)--(r2y);

\node[
    font=\small,
    align=left,
    anchor=west
] at (1.05,-0.75)
{open\\connectors};

\draw[
    ->,
    >=stealth,
    orange!85!yellow,
    line width=1.0pt
]
(2.15,-0.6)
to[out=-20]
(3.10,-1.00);

\draw[
    ->,
    >=stealth,
    orange!85!yellow,
    line width=1.0pt
]
(1.75,-0.3)
to[in=10]
(0.80,0.2);

\node[
    font=\small,
    align=left,
    anchor=west
] at (5.75,0.55)
{ghost\\connector};

\draw[
    ->,
    >=stealth,
    orange!85!yellow,
    line width=1.0pt
]
(6.35,0.15)
to[out=-40,in=30]
(6.05,-0.70);

\end{tikzpicture}}
\caption{
Illustration of three routers $\rho_1,\rho_2,\rho_3$ and six connectors
$\kappa_1,\ldots,\kappa_6$ relative to a large vertex set $C$ satisfying the invariants. The pale red regions represent routers, with their skeletons highlighted
in red. The pale blue regions represent connectors, whose centers are
the large hollow vertices. The green region represents the ghost
connector $\kappa_6$.  Router $\rho_2$ is pending, and routers $\rho_1$ and $\rho_3$ are ready.  Orange vertices certify invariants 5 and 6 of their corresponding supernodes. }
\label{fig:routers-connectors-C}
\end{figure}

In every step, we maintain a tuple $(\cS,\cX, I, C)$ satisfying the following invariants:

\begin{tcolorbox}
    \textbf{\hypertarget{InvsSXIC}{Invariants of $(\cS, \cX,I, C)$}}:
     \begin{enumerate}
     \item  $\cS$ is a set of routers, $\cX$ is a set of only connectors. We denote by $\cX_o, \cX_c, \cX_g$ the sets of open, closed, and ghost connectors of $\cX$, respectively. So $(\cX_o, \cX_c, \cX_g)$ is a partition of $\cX$. Every open/ghost connector in $\cX$ is associated with one router in $\cS$. A router that has no associated open connector in $\cX$ is called a \EMPH{pending router}; otherwise, it is called a \EMPH{ready router}. A ready router is associated with exactly one open connector, but may be associated with multiple ghost connectors.
     Each pending router must acquire an open connector before the next router can be created. Once every router is ready, we can create a new router and use the open connectors to connect it to the existing routers.
     \item  $I$ is a set of vertices of $G$. 
        \item $C$ is a connected component of $G - I - \bigcup _{\eta \in \cS\cup \cX} \eta^+$ of size at least $2n/3$. By definition of $\eta^+$, we have that $C$ is not adjacent to the skeleton of any supernode in $\cS\cup \cX$.
        \item  $\cX_c\cup \cS$ forms an induced model of $\dot K_{h}$ for  $h=|\cS|\leq t-1$.
        \item  For every  open or ghost connector $\kappa\in \cX_o \cup \cX_g$, we have $(\kappa\cap N_G(C)) \setminus (\cS\cup \cX \setminus \{\kappa\})^+ \not= \varnothing$. That is, at least one vertex adjacent to $C$ in $\kappa$   is not guarded by other supernodes.
        \item  For every pending router $\rho\in \cS$, $\rho$ has no associated ghost connector and     $(\rho\cap N_G(C)) \setminus (\cS \cup \cX \setminus \{\rho\})^+ \not= \varnothing$.
    \end{enumerate}
See \Cref{fig:routers-connectors-C} for an illustration. 
\end{tcolorbox}

Our final separator will consist of $I$ vertices (as singletons), and stars around the vertices in the skeletons of all the supernodes in $\cS\cup\cX$.
Intuitively, invariant 3 means that $C$ is the largest connected component if we remove the ``current separator'' from $G$. 

\paragraph{The construction.} At the outset, we pick an arbitrary vertex $v \in V(G)$, and form a supernode $\eta_{v} = N_G[v]$. Let $C$ be the largest connected component of $G\setminus \eta_v$. If $|C|\leq 2n/3$, then $\cF = \{\eta_v\}$ is a star-based separator of size $1$. Otherwise,  $(\cS \coloneqq \{\eta_{v}\}, \cX \coloneqq \varnothing, I \coloneqq \varnothing, C)$ is the tuple satisfying all the 6 invariants for the next step. 

Now consider the inductive step with the tuple  $(\cS, \cX, I, C)$.  We say that $\cS$ is \EMPH{complete} if all the routers in $\cS$ are ready. 
When $\cS$ is complete, we can extend the induced model by
constructing a new router using vertices of $C$ together
with one suitably chosen vertex from each open connector,
as described in Case 2a below.

\begin{figure}[!htb]
\centering
\resizebox{\textwidth}{!}{\begin{tikzpicture}[
    x=1cm,
    y=1cm,
    line cap=round,
    line join=round,
    ambientv/.style={
        circle,
        fill=black,
        inner sep=1.8pt
    },
    orangev/.style={
        circle,
        fill=orange!85!yellow,
        inner sep=2.2pt
    },
    hollowv/.style={
        circle,
        draw=black,
        fill=white,
        line width=1.2pt,
        minimum size=5.4mm,
        inner sep=0pt
    },
    skedge/.style={
        red!85!black,
        line width=1.7pt
    },
    ambientedge/.style={
        black,
        line width=.75pt
    },
    modeledge/.style={
        blue!75!cyan,
        line width=2.4pt
    },
    routerlabel/.style={
        text=red!80!black,
        font=\Large
    },
    connectorlabel/.style={
        text=blue!75!black,
        font=\Large
    },
    treeedge/.style={
        draw=black,
        line width=.9pt
    }
]

\begin{scope}[on background layer]
    \fill[red!9]
        (-6.20,2.45) rectangle (-4.35,4.55);

    \begin{scope}[shift={(-7.20,2.70)},rotate=12]
        \fill[blue!9]
            (0,0) ellipse [x radius=.48cm,y radius=.72cm];
    \end{scope}
\end{scope}

\node[ambientv] (at-rx1) at (-5.80,4.28) {};

\node[hollowv] (at-r1) at (-5.25,3.90) {};
\node[hollowv] (at-r2) at (-5.75,3.55) {};
\node[hollowv] (at-r3) at (-4.85,3.45) {};
\node[hollowv] (at-r4) at (-5.45,2.90) {};

\node[ambientv] (at-rx2) at (-6.10,2.55) {};
\node[ambientv] (at-rx3) at (-4.85,2.65) {};

\draw[ambientedge] (at-rx1)--(at-r1);
\draw[ambientedge] (at-rx1)--(at-r2);

\draw[skedge] (at-r2)--(at-r1);
\draw[skedge] (at-r1)--(at-r4);
\draw[skedge] (at-r4)--(at-r3);

\draw[ambientedge] (at-r4)--(at-rx2);
\draw[ambientedge] (at-r4)--(at-rx3);

\node[routerlabel] at (-3.95,4.15) {$\rho$};

\node[hollowv] (at-kc) at (-7.20,2.80) {};
\node[ambientv] (at-ka) at (-7.35,3.35) {};
\node[ambientv] (at-kb) at (-6.90,2.25) {};

\draw[ambientedge] (at-ka)--(at-kc);
\draw[ambientedge] (at-kc)--(at-kb);

\draw[modeledge] (-8.50,2.85)--(at-kc);
\draw[modeledge] (at-kc)--(at-r2);

\node[orangev] (at-x) at (-5.02,2.32) {};
\node[text=blue!75!black,font=\small,anchor=east] at (-5.17,2.30) {$x$};

\draw[ambientedge] (at-r4)--(at-x);

\node[ambientv] (at-c1) at (-5.45,1.55) {};
\node[ambientv] (at-c2) at (-4.75,1.55) {};
\node[ambientv] (at-c3) at (-4.20,1.70) {};

\draw[ambientedge] (at-x)--(at-c1);
\draw[ambientedge] (at-x)--(at-c2);
\draw[ambientedge] (at-rx3)--(at-c3);

\draw[black,line width=.8pt]
(-6.50,1.18)
to[out=45,in=175]
(-5.25,1.95)
to[out=-5,in=150]
(-3.55,1.55);

\node[font=\Large] at (-5.80,1.18) {$C$};

\draw[
    ->,
    >=stealth,
    orange!85!yellow,
    line width=1.1pt
]
(-7.25,1.05)
to[out=-120,in=120]
(-7.25,0.15);

\begin{scope}[on background layer]
    \fill[red!9]
        (-6.15,-1.95) rectangle (-4.50,0.20);

\begin{scope}[shift={(-7.35,-1.85)},rotate=8]
        \fill[blue!9]
            (0,0) ellipse [x radius=.47cm,y radius=.70cm];
    \end{scope}

\begin{scope}[shift={(-5.15,-2.35)},rotate=-3]
        \fill[blue!9]
            (0,0) ellipse [x radius=.53cm,y radius=.76cm];
    \end{scope}
\end{scope}

\node[ambientv] (ab-rx1) at (-5.85,-0.05) {};

\node[hollowv] (ab-r1) at (-5.28,-0.42) {};
\node[hollowv] (ab-r2) at (-5.82,-0.80) {};
\node[hollowv] (ab-r3) at (-5.38,-1.35) {};

\node[ambientv] (ab-rx2) at (-6.10,-1.82) {};
\node[ambientv] (ab-rx3) at (-4.75,-1.72) {};

\draw[ambientedge] (ab-rx1)--(ab-r1);
\draw[ambientedge] (ab-rx1)--(ab-r2);

\draw[skedge] (ab-r2)--(ab-r1);
\draw[skedge] (ab-r1)--(ab-r3);

\draw[ambientedge] (ab-r3)--(ab-rx2);
\draw[ambientedge] (ab-r3)--(ab-rx3);

\node[routerlabel] at (-4.20,-0.35) {$\rho$};

\node[hollowv] (ab-k0) at (-7.35,-1.75) {};
\node[ambientv] (ab-k0a) at (-7.48,-1.20) {};
\node[ambientv] (ab-k0b) at (-7.05,-2.27) {};

\draw[ambientedge] (ab-k0a)--(ab-k0);
\draw[ambientedge] (ab-k0)--(ab-k0b);

\draw[modeledge] (-8.60,-1.70)--(ab-k0);
\draw[modeledge] (ab-k0)--(ab-r2);

\node[hollowv] (ab-kc) at (-5.15,-2.10) {};
\node[ambientv] (ab-ka) at (-5.48,-2.80) {};
\node[ambientv] (ab-kb) at (-4.87,-2.82) {};

\draw[ambientedge] (ab-kc)--(ab-ka);
\draw[ambientedge] (ab-kc)--(ab-kb);

\draw[modeledge] (ab-r3)--(ab-kc);

\node[text=blue!75!black,font=\small,anchor=east] at (-4.90,-2.1) {$x$};
\node[connectorlabel] at (-6.00,-2.35) {$\kappa$};

\node[ambientv] (ab-out) at (-4.25,-2.65) {};
\draw[ambientedge] (ab-rx3)--(ab-out);

\node[ambientv] (ab-c1) at (-5.75,-3.55) {};
\node[ambientv] (ab-c2) at (-5.20,-3.60) {};

\draw[ambientedge] (ab-ka)--(ab-c1);
\draw[ambientedge] (ab-kb)--(ab-c2);

\draw[black,line width=.8pt]
(-7.15,-3.90)
to[out=55,in=175]
(-6.15,-3.28)
to[out=-3,in=178]
(-5.35,-3.14)
to[out=-2,in=178]
(-4.78,-3.06)
to[out=14,in=168]
(-4.08,-2.40)
to[out=-22,in=105]
(-3.78,-3.55);

\node[font=\Large] at (-6.20,-3.55) {$C$};

\node[font=\Large] at (-5.35,-4.40) {(a)};

\begin{scope}[on background layer]
\filldraw[
    fill=gray!18,
    draw=black,
    line width=.7pt
]
plot[smooth cycle,tension=.85] coordinates {
    (0.35,-1.05)
    (0.75,-0.15)
    (1.80,0.10)
    (2.65,-0.05)
    (3.25,-0.65)
    (4.40,-0.85)
    (5.55,-0.78)
    (6.55,-0.45)
    (7.85,-0.05)
    (8.95,-0.35)
    (9.45,-1.45)
    (8.85,-2.55)
    (7.10,-3.30)
    (4.75,-3.55)
    (2.50,-3.30)
    (0.75,-2.45)
};
\end{scope}

\node[font=\Huge] at (3.35,-2.65) {$C$};

\begin{scope}[on background layer]
\fill[red!9] (1.55,2.25) rectangle (3.00,4.35);

\fill[red!9] (4.30,0.75) rectangle (5.45,2.25);

\fill[red!9] (7.70,2.10) rectangle (9.25,4.45);
\end{scope}

\begin{scope}[on background layer]
\begin{scope}[shift={(1.85,1.10)},rotate=12]
        \fill[blue!9] (0,0) ellipse [x radius=.58cm,y radius=.92cm];
    \end{scope}

\begin{scope}[shift={(4.85,0)},rotate=-8]
        \fill[blue!9] (0,0) ellipse [x radius=.62cm,y radius=.70cm];
    \end{scope}

\begin{scope}[shift={(8.25,0.90)},rotate=-10]
        \fill[blue!9] (0,0) ellipse [x radius=.66cm,y radius=.85cm];
    \end{scope}

\begin{scope}[shift={(5.28,3.90)},rotate=0]
        \fill[blue!9] (0,0) ellipse [x radius=.70cm,y radius=1.05cm];
    \end{scope}

\begin{scope}[shift={(3.55,2.25)},rotate=-28]
        \fill[blue!9] (0,0) ellipse [x radius=.60cm,y radius=.48cm];
    \end{scope}

\begin{scope}[shift={(6.35,2.40)},rotate=-15]
        \fill[blue!9] (0,0) ellipse [x radius=.58cm,y radius=.78cm];
    \end{scope}
\end{scope}

\node[ambientv] (b-r1x1) at (1.90,3.72) {};

\node[hollowv] (b-r1a) at (2.38,3.95) {};
\node[hollowv] (b-r1b) at (1.92,3.12) {};
\node[hollowv] (b-r1c) at (2.58,3.32) {};
\node[hollowv] (b-r1d) at (2.48,2.60) {};

\node[ambientv] (b-r1x2) at (2.08,2.55) {};

\draw[ambientedge] (b-r1x1)--(b-r1a);
\draw[skedge] (b-r1b)--(b-r1a);
\draw[skedge] (b-r1a)--(b-r1c);
\draw[skedge] (b-r1c)--(b-r1d);
\draw[ambientedge] (b-r1x2)--(b-r1c);
\draw[ambientedge] (b-r1x2)--(b-r1d);

\node[routerlabel] at (3.28,4.15) {$\rho_1$};

\node[ambientv] (b-r2x1) at (8.12,4.08) {};

\node[hollowv] (b-r2a) at (8.56,3.75) {};
\node[hollowv] (b-r2b) at (8.02,3.35) {};
\node[hollowv] (b-r2c) at (8.82,3.05) {};
\node[hollowv] (b-r2d) at (8.40,2.52) {};

\node[ambientv] (b-r2x2) at (8.22,2.18) {};

\draw[ambientedge] (b-r2x1)--(b-r2a);
\draw[ambientedge] (b-r2x1)--(b-r2b);

\draw[skedge] (b-r2b)--(b-r2a);
\draw[skedge] (b-r2a)--(b-r2d);
\draw[skedge] (b-r2d)--(b-r2c);

\draw[ambientedge] (b-r2d)--(b-r2x2);

\node[routerlabel] at (9.62,4.12) {$\rho_2$};

\node[hollowv] (b-r3a) at (5.05,1.90) {};
\node[hollowv] (b-r3b) at (4.53,1.43) {};
\node[hollowv] (b-r3c) at (5.10,1.13) {};

\node[ambientv] (b-r3x) at (4.62,0.98) {};

\draw[skedge] (b-r3a)--(b-r3b);
\draw[skedge] (b-r3a)--(b-r3c);
\draw[ambientedge] (b-r3x)--(b-r3b);
\draw[ambientedge] (b-r3x)--(b-r3c);

\node[routerlabel] at (4.72,2.60) {$\rho_3$};

\node[hollowv] (b-k4c) at (5.28,3.82) {};

\node[ambientv] (b-k4a) at (5.02,4.50) {};
\node[ambientv] (b-k4b) at (5.65,4.50) {};
\node[ambientv] (b-k4d) at (4.98,3.25) {};
\node[ambientv] (b-k4e) at (5.40,2.98) {};

\draw[ambientedge] (b-k4c)--(b-k4a);
\draw[ambientedge] (b-k4c)--(b-k4b);
\draw[ambientedge] (b-k4c)--(b-k4d);
\draw[ambientedge] (b-k4c)--(b-k4e);

\node[connectorlabel] at (4.15,4.45) {$\kappa_4$};

\node[hollowv] (b-k5c) at (3.62,2.35) {};
\node[ambientv] (b-k5a) at (3.30,1.93) {};

\draw[ambientedge] (b-k5c)--(b-k5a);

\node[connectorlabel] at (3.50,3.10) {$\kappa_5$};

\node[hollowv] (b-k6c) at (6.35,2.55) {};
\node[ambientv] (b-k6a) at (6.18,3.20) {};
\node[ambientv] (b-k6b) at (6.70,1.92) {};

\draw[ambientedge] (b-k6a)--(b-k6c);
\draw[ambientedge] (b-k6c)--(b-k6b);

\node[connectorlabel] at (6.92,3.25) {$\kappa_6$};

\node[hollowv] (b-k1c) at (1.92,1.40) {};

\node[ambientv] (b-k1a) at (1.68,1.98) {};
\node[ambientv] (b-k1b) at (2.02,0.72) {};
\node[orangev]  (b-y1)  at (1.52,0.98) {};

\draw[ambientedge] (b-k1a)--(b-k1c);
\draw[ambientedge] (b-k1c)--(b-k1b);
\draw[ambientedge] (b-k1c)--(b-y1);

\node[connectorlabel] at (1.05,1.80) {$\kappa_1$};

\node[text=blue!75!black,font=\small,anchor=east] at (1.42,0.86) {$y_1$};

\node[hollowv] (b-k2c) at (4.92,0.38) {};

\node[ambientv] (b-k2a) at (4.50,-0.05) {};
\node[orangev]  (b-k2o) at (5.32,0.05) {};
\node[orangev]  (b-y2)  at (5.02,-0.48) {};

\draw[ambientedge] (b-k2c)--(b-k2a);
\draw[ambientedge] (b-k2c)--(b-k2o);
\draw[ambientedge] (b-k2c)--(b-y2);

\node[connectorlabel] at (5.92,0.30) {$\kappa_2$};

\node[text=blue!75!black,font=\small,anchor=west] at (5.18,-0.55) {$y_2$};

\node[hollowv] (b-k3c) at (8.25,1.35) {};

\node[ambientv] (b-k3a) at (7.88,0.93) {};
\node[orangev]  (b-y3)  at (8.35,0.53) {};

\draw[ambientedge] (b-k3c)--(b-k3a);
\draw[ambientedge] (b-k3c)--(b-y3);

\node[connectorlabel] at (8.95,1.25) {$\kappa_3$};

\node[text=blue!75!black,font=\small,anchor=west] at (8.48,0.35) {$y_3$};

\draw[modeledge] (b-r1c)--(b-k4c);
\draw[modeledge] (b-k4c)--(b-r2b);

\draw[modeledge] (b-r1d)--(b-k5c);
\draw[modeledge] (b-k5c)--(b-r3b);

\draw[modeledge] (b-r2b)--(b-k6c);
\draw[modeledge] (b-k6c)--(b-r3a);

\draw[modeledge] (b-r1d)--(b-k1c);
\draw[modeledge] (b-r3c)--(b-k2c);
\draw[modeledge] (b-r2x2)--(b-k3c);

\draw[ambientedge] (b-k1b)--(b-r3b);
\draw[ambientedge] (b-k5c)--(b-k2a);
\draw[ambientedge] (b-k6c)--(b-r2x2);
\draw[ambientedge] (b-r3c)--(b-k3a);

\coordinate (z1pos) at (2.02,-0.82);
\coordinate (z2pos) at (4.62,-1.38);
\coordinate (z3pos) at (8.28,-0.57);
\coordinate (apos)  at (4.72,-2.15);

\begin{scope}[on background layer]

\draw[green!35,line width=5pt,line cap=round]
(b-y1)--(z1pos);

\draw[green!35,line width=5pt,line cap=round]
(b-y2)--(z2pos);

\draw[green!35,line width=5pt,line cap=round]
(b-y3)--(z3pos);

\draw[green!35,line width=5pt,line cap=round]
(z1pos)
to[out=-5,in=165]
(3.05,-1.55)
to[out=-15,in=170]
(4.45,-1.62);

\draw[green!35,line width=5pt,line cap=round]
(z2pos)
to[out=-80,in=80]
(4.50,-1.72);

\draw[green!35,line width=5pt,line cap=round]
(4.45,-1.62)
to[out=-30,in=90]
(apos);

\draw[green!35,line width=5pt,line cap=round]
(z3pos)
to[out=-85,in=10]
(7.85,-1.40)
to[out=190,in=5]
(5.20,-1.55)
to[out=185,in=20]
(4.45,-1.62);

\end{scope}

\node[ambientv] (z1) at (z1pos) {};
\node[ambientv] (z2) at (z2pos) {};
\node[ambientv] (z3) at (z3pos) {};
\node[ambientv] (aa) at (apos) {};

\node[text=blue!75!black,font=\small,anchor=north] at (1.88,-0.95) {$z_1$};
\node[text=blue!75!black,font=\small,anchor=west] at (4.78,-1.35) {$z_2$};
\node[text=blue!75!black,font=\small,anchor=west] at (8.43,-0.55) {$z_3$};
\node[font=\small,anchor=west] at (4.90,-2.28) {$v$};

\draw[ambientedge] (b-k1b)--(z1);
\draw[ambientedge] (b-y1)--(z1);

\draw[ambientedge] (b-k2a)--(z2);
\draw[ambientedge] (b-y2)--(z2);

\draw[ambientedge] (b-y3)--(z3);

\draw[treeedge]
(z1)
to[out=-5,in=165]
(3.05,-1.55)
to[out=-15,in=170]
(4.45,-1.62);

\draw[treeedge]
(z2)
to[out=-80,in=80]
(4.50,-1.62);

\draw[treeedge]
(4.45,-1.62)
to[out=-30,in=90]
(aa);

\draw[treeedge]
(z3)
to[out=-85,in=10]
(7.85,-1.40)
to[out=190,in=5]
(5.20,-1.55)
to[out=185,in=20]
(4.45,-1.62);

\node[text=red!80!black,font=\Large] at (7.10,-2.05) {$T_\rho$};

\node[font=\Large] at (5.00,-4.20) {(b)};

\end{tikzpicture}}
\caption{
Schematic illustration of the router and connector constructions.
Subfigure (a) illustrates the construction of connector $\kappa$ in Case 1. Subfigure (b) illustrates the skeleton $T_\rho$ (the green highlighted paths) of the new router $\rho$ constructed in Subcase 2a 
inside $C$.}
\label{fig:router-connector-tree}
\end{figure}
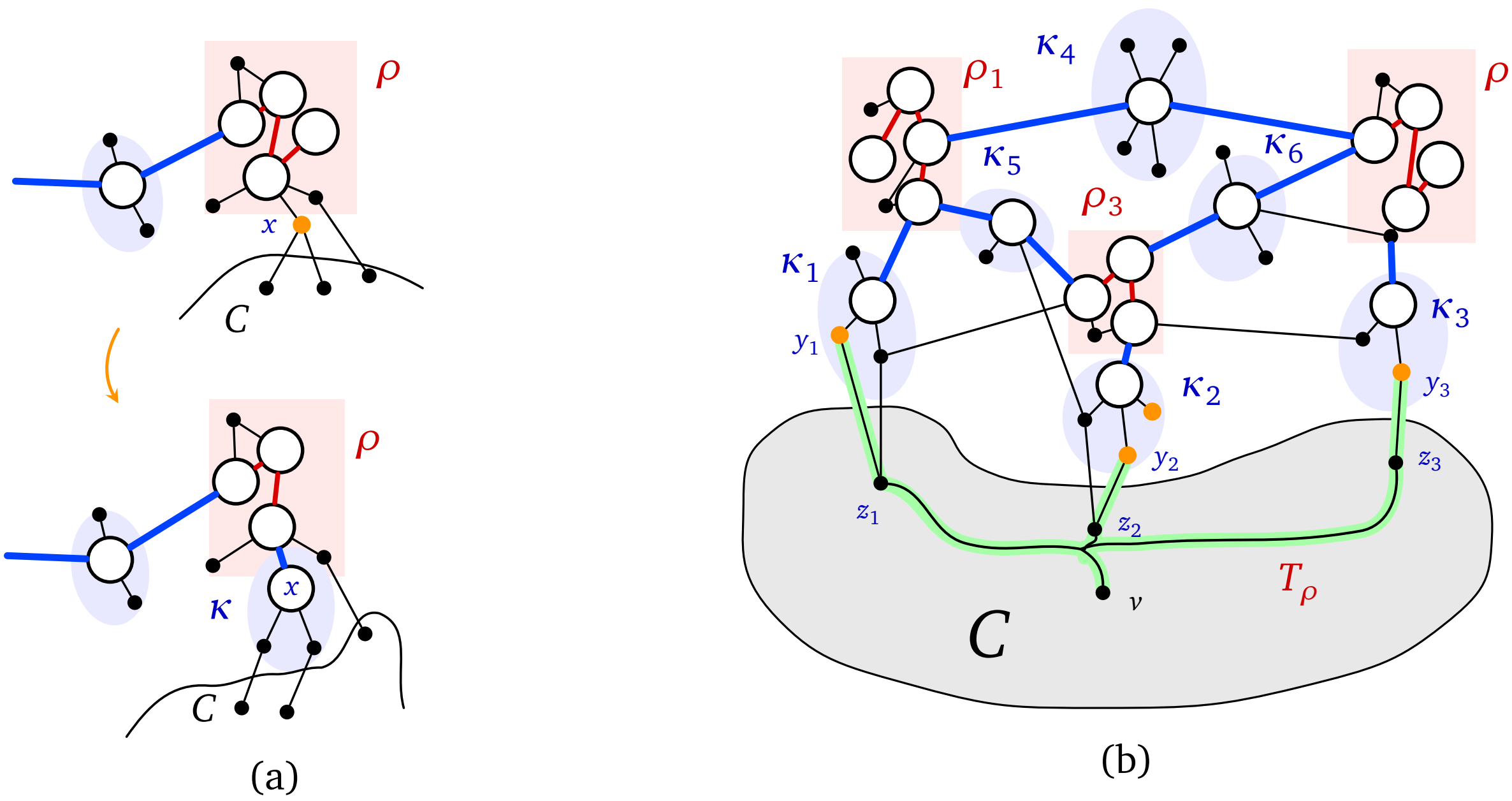

\subsection{Creating new routers/connectors}
We consider two cases:

\paragraph{[Case 1: $\cS$ is not complete, creating an open connector].}\hypertarget{Case1}{} Let $\rho\in \cS$ be a pending router. Invariant 6 implies that there exists a vertex $x\in \rho \cap N_G(C)$ such that $x$ is not adjacent to any skeleton of any supernode in $\cS \cup \cX \setminus \{\rho\}$. We construct a new open connector $\kappa=N_{G[C\cup\{x\}]}[x]$ with skeleton $T_\kappa=\{x\}$ that contains $x$ and all the neighbors of $x$ in $C$, and add $\kappa$ to $\cX_o$ (and $\cX$). See \Cref{fig:router-connector-tree}(a).
$C$ is updated to be the largest connected component of $C-\kappa$.

\paragraph{[Case 2: $\cS$ is complete.]}  Pick an arbitrary vertex $v\in C$. For all $i \geqslant 0$, let $L_{i}:= \{w \in V(C) : \dist_{C}(v, w) = i\}$; so $(L_0, L_1, \dots)$ is a \EMPH{BFS layering} of $C$ \EMPH{rooted} at $v$. Every non-empty set $L_{i}$ is called a \EMPH{layer}. Note that $L_0 = \{v\}$. Associated with a BFS layering is a \EMPH{BFS spanning tree} $T_v$ obtained by choosing, for each non-root vertex $w \in L_i$ with $i \geqslant 1$, a neighbor $u$ in $L_{i - 1}$, and adding the edge $wu$ to $T_v$.
We consider two subcases:

\begin{itemize}
\item  \textbf{[Subcase 2a:  $T_v$ has at most $2\ell \log(n) = O(\ell \log n)$ layers, creating a router].~} By invariant 1 and since $\cS$ is complete, each router in $\cS$ is associated with exactly one open connector in $\cX_o$ and hence $|\cS| = |\cX_o|$.  For every open connector $\kappa \in \cX_o$, using invariant 5, let $y_{\kappa} \in (\kappa\cap N_G(C)) \setminus (\cS\cup \cX \setminus \{\kappa\})^+$ be a vertex in $\kappa$ adjacent to a vertex  $z_\kappa\in C$ (and not covered by other supernodes). 
By invariant 3, $y_{\kappa} \notin T_{\kappa}$. 
Define a new skeleton $T_{\rho}$ 
by taking the subtree of $T_v$ consisting of $|\cX_o|$ shortest paths from $v$ towards $\{z_\kappa\}_{\kappa\in\cX_o}$, and in addition the edges $\{\{y_\kappa,z_\kappa\}\}_{\kappa\in\cX_o}$.
See \Cref{fig:router-connector-tree}(b). Then we create a new router $\rho$ (with skeleton $T_\rho$) by adding to $T_\rho$ all the neighbors of
$T_\rho$ in $C$ (that is $\rho=T_\rho\cup (N_{G}[T_{\rho}]\cap C)$). $\rho$ is added to~$\cS$. Then we update $\cX_c \leftarrow \cX_c \cup \cX_o$ and set $\cX_o = \varnothing$, as all connectors in $\cX_o$ are now closed. 
$C$ is updated to be the largest connected component of $C-\rho$.

    \item  \textbf{[Subcase 2b:  $T_v$ has more than $2\ell \log(n) = O(\ell \log n)$ layers].~} In this case, we  add more vertices to $I$.

For every $i \geqslant 0$, let $L_{\geq i} := L_{i} \cup L_{i + 1} \cup \dots$, and let $L_{> i} := L_{\geq i} \setminus L_i = L_{\geq i + 1}$. For every $i \geqslant 0$, let $L_{\leq i} := L_{0} \cup L_{1} \cup \dots \cup L_{i}$, and let $L_{< i} := L_{\leq i} \setminus L_i = L_{\leq i - 1}$.

\begin{claim}\label{clm:sep-layer} There exists $i \geqslant 1$ such that $0 < |L_i| \leq  \min\{|L_{> i}|, |L_{< i}|\}/\ell$.
\end{claim}
\begin{proof}
Color a layer $L_j$ \EMPH{red} if $|L_{j}| > |L_{< j}|/\ell$ and color $L_j$ \EMPH{blue} if $|L_{j}| >  |L_{> j}|/\ell$. (One layer could have both colors.) Observe that the number of red layers is at most $\ell \log (n)$. This is because every time we encounter a red layer $j$, the number of vertices in $L_{\leq j}$ increases by a factor of $(1+1/\ell)$, and hence the total number of vertices of $G$ when seeing the $(\ell \log n)$-th red layer is at least:
\begin{equation*}
    (1+1/\ell)^{\ell \log(n)} \geq 2^{\log(n)} = n,
\end{equation*}
and hence the number of red layers is at most $\ell \log (n)$. By a similar argument, the number of blue layers is at most $\ell \log (n)$. Thus, there must be an uncolored layer, which gives $L_i$.
\end{proof}

We add $L_i$ to $I$ and \EMPH{charge} the size of $L_i$ to vertices NOT in the largest connected component of $C-L_i$. By \Cref{clm:sep-layer}, each vertex is charged an amount of at most $1/\ell$. Then we update $C$ to be the largest connected component of $C-L_i$. This implies that every vertex will be charged at most once during the entire process.  
\end{itemize}

This finishes the creation step.
If $|C|\geq 2n/3$ (after the update), then we proceed to the cleaning step below. 
Otherwise, we have found our separator \EMPH{$\cF = \cX\cup \textsc{BreakStar}(\cS) \cup I$}, where $\EMPH{\textsc{BreakStar}}(\cS)$ breaks every router $\rho \in \cS$ into stars and collects all the resulting stars. 
More specifically, $\rho$ is broken into stars by taking each vertex $v$ in its skeleton $T_{\rho}$ as the center of its star, and assigning each vertex $u\in \rho - T_{\rho}$ to the star of its (arbitrary) neighbor in $T_{\rho}$.

\subsection{Cleaning step}

In all cases in the creation step, $C$ is shrunk. 

\begin{observation}\label{obs:model}  In Case 1, the skeleton of the new connector $\kappa$ is not adjacent to the skeleton of any supernode in $\cS\cup \cX$ other than its associated router.  In Case 2a, after adding $\rho$ to $\cS$, $\cX_c\cup \cS$ is an induced model of size $|\cS|$.
\end{observation}
\begin{proof}    
    The first claim about $\kappa$ holds by the choice of its center $x$.
    For the second claim, invariant 3 implies that the vertices
    of $T_\rho$ in $C$ have no neighbors in the existing skeletons.
    For each $\kappa\in\cX_o$, the vertex $y_\kappa$ is adjacent
    to the center of $\kappa$, and invariant 5 guarantees that
    it is not adjacent to any other existing skeleton.
    Thus $T_\rho$ is adjacent precisely to the centers of the
    open connectors. These connectors become closed, while
    all previous skeleton adjacencies are preserved, yielding
    the required induced model.
\end{proof}

\paragraph{Cleaning $(\cS, \cX,I,C)$.}  The goal of this cleaning step is to restore the 6 invariants of  $(\cS, \cX,I,C)$. The two key invariants are 5 and 6. In this step, we iteratively apply cleaning open/ghost connectors and cleaning pending routers until there is no violation.
\begin{tcolorbox}[breakable]
 \textbf{\hypertarget{CleaningStep}{Cleaning $\cS\cup \cX$}}: Iteratively apply the following steps until invariants 5 and 6 hold:
\begin{enumerate}
    \item \textbf{Cleaning open/ghost connectors.~} There are two ways that an open/ghost connector $\kappa \in \cX_o\cup \cX_g$ violates invariant 5 w.r.t. $C$: 
\begin{itemize}
    \item (i)  $\kappa$ has no neighbor in $C$. In this case, we simply remove $\kappa$ from $\cX$.
    \item (ii) $\kappa\cap N_G(C) \subseteq (\cS\cup \cX  \setminus \{\kappa\})^+$. In this case, for every $v\in \kappa\cap N_G(C)$, let $\eta\in \cS\cup \cX \setminus \{\kappa\}$ be an arbitrary supernode such that $v\in \eta^+$. We then reassign $v$ to $\eta$ and remove $v$ from  $\kappa$. Finally, we remove $\kappa$ from $\cX$. 
\end{itemize}
We greedily clean each open/ghost connector $\kappa \in \cX$ until every open/ghost connector satisfies invariant 5. Then we move to the second step. 
\item \textbf{Transforming ghost routers to ready routers.~}   This step applies when a pending router $\rho \in \cS$ is associated with at least one ghost connector in $\cX_g$.
(Invariant 6 says that a pending router has no associated ghost connector.) Such a router is called a \EMPH{ghost router}.  
One way this could happen is that after we clean open/ghost connectors in $\cX$ in step 1, some ready router $\rho \in \cS$ becomes pending when it loses the associated open connector. 
Since $\rho$ was a ready router before it was made pending, it may still have (possibly more than one) associated ghost connectors. 
Then, for each ghost router $\rho$, we simply designate one of its ghost connectors as an open connector (and effectively move this connector from $\cX_g$ to $\cX_o$), making $\rho$ a ready router. 
Once all the ghost routers are converted to ready, we proceed to the next step.
\item  \textbf{Cleaning pending routers.~} At this point, each pending router has no associated ghost connectors.   Note that step 1 might make some ready routers in $\cS$ become pending. 
There are still two ways a pending router $\rho$ violates invariant 6 w.r.t. $C$:
\begin{itemize}
    \item (i)  $\rho$ has no neighbor in $C$. In this case, we remove $\rho$ from $\cS$.
    \item (ii) $\rho\cap N_G(C) \subseteq (\cS\cup \cX  \setminus \{\rho\})^+$. 
In this case, for every $v\in \rho\cap N_G(C)$, let $\eta\in \cS\cup \cX \setminus \{\rho\}$ be an arbitrary supernode such that $v\in \eta^+$. We reassign $v$ to $\eta$.
     Then, we remove $\rho$ from $\cS$.  
\end{itemize}
Once $\rho$ is removed from $\cS$, we mark all its associated closed connectors in $\cX_c$ as ghost connectors. (Basically, we reassign them from $\cX_c$ to $\cX_g$.)

Note that the new ghost connectors might violate invariant 5. Furthermore, it might happen that some pending router becomes associated with one of these new ghost connectors, violating invariant 6.
Hence we process at most a single pending router in this step.
After removing $\rho$ and relabeling its associated closed
connectors, before we look for additional pending routers that violate invariant 6, we return to the beginning of the cleaning procedure, and apply the 3 steps again.

\end{enumerate}

\end{tcolorbox}

The above cleaning process will eventually stop because only Step 3 creates ghost connectors, which could result in another iteration. However, at each Step 3, one (pending) router is removed from $\cS$. By invariant 4, $\cS$ has at most $t-1$ routers, so the number of iterations is at most $t-1$.

 Let  $(\cS,\cX,I,C)$ be the tuple after the cleaning.  The cleaning process directly guarantees invariants 5 and 6.   Note that the cleaning process does not change $C$, and hence invariants 1 and 2 remain true for $(\cS,\cX,I,C)$. We now show that invariant 3 holds. Since the cleaning step removes supernodes from $\cS \cup \cX$, the set $\bigcup_{\eta \in \cS \cup \cX}\eta^+$ might shrink; and so the fact that $C$ is unchanged does not directly by itself imply that $C$ remains a connected component of $G - I - \bigcup_{\eta \in \cS \cup \cX}\eta^+$. However, the skeletons of surviving supernodes are unchanged, and so $C$ remains disjoint from $\bigcup_{\eta \in \cS \cup \cX}\eta^+$. By \Cref{lm:cF-separate} (which we prove below), every vertex in $N_{G}(C) \setminus C$ remains in $I$ or in the supernodes of $\cS \cup \cX$, and so $N_{G}(C) \setminus C \subseteq I \cup \bigcup_{\eta \in \cS \cup \cX}\eta^+$. So $C$ remains to be a connected component of $G - I - \bigcup_{\eta \in \cS \cup \cX}\eta^+$, and invariant 3 is satisfied.
 
 Invariant 4 holds because if we remove a router $\rho$ from $\cS$, we relabel all its associated closed connectors as ghost connectors (and put them in $\cX_g$). Thus, $\cS\cup \cX_c$ remains an induced model of $\dot K_{h}$ for some $h$. As $G$ excludes $\dot K_t$ as an induced minor, $h\leq t-1$.  Thus, $(\cS,\cX,I,C)$ satisfies all the 6 invariants. And the next iteration can continue processing  $(\cS,\cX,I,C)$, until we find our separator $\cF$.

 \subsection{Bounding the size of $\cF$.} 
 Before bounding the size of $\cF$, we bound the number of ghost connectors. By invariants 1 and 4, $|\cX_c| \leq {|\cS| \choose 2} \leq (t-1)^2$ and $|\cX_o| \leq |\cS| \leq t-1$, but it is not immediately clear that $|\cX_g|$ is bounded.

 \begin{lemma}\label{lm:ghost-bound} At all times, $|\cX|\leq   t(t-1)$ (which trivially implies that $|\cX_g|\leq t(t-1)$).
 \end{lemma}
 \begin{proof}
   For every router $\rho\in \cS$, let $\cX(\rho)$ (and $\cX_c(\rho)$, $\cX_o(\rho)$, $\cX_g(\rho)$, respectively) be the set of (closed, open, ghost, respectively) connectors in $\cX$ associated with $\rho$. 
   We prove a stronger claim: for every router $\rho \in \cS$, $|\cX(\rho)|\leq t$.  
   
   Observe that $|\cX_c(\rho)| \leq t-1$ by invariant 4.   If $\rho$ is a pending router, then by invariant 6 we have $\cX_g(\rho) = \emptyset$, and so by invariants 1 and 4, we have $\cX(\rho) = \cX_c(\rho)$, giving $|\cX(\rho)| \leq t-1$.  
   
   Herein, we consider the case where $\rho$ is a ready router. Then $|\cX_o(\rho)| = 1$.   Observe that cleaning does not create a new connector for $\rho$: it either only removes or relabels the connectors of $\cX(\rho)$. Also, $\rho$ might get a new open connector in Case 1, but the condition is that $\rho$ must be a pending router, which has at most $t-1$ closed connectors (and no ghost connectors) and thus $|\cX(\rho)| \leq t$. If we create a new router (in Subcase 2a),  the (only) open connector of $\cX_o(\rho)$  is marked closed, and hence moved to $\cX_c(\rho)$, and therefore, the total number of connectors in $\cX(\rho)$ does not change. Thus, in every case, $|\cX(\rho)|\leq t$, giving  $|\cX| \leq t |\cS| \leq t(t-1)$. 
 \end{proof}

 Now we bound the size of $\cF$.  By \Cref{clm:sep-layer}, and the fact that each vertex is charged for $I$ at most once during the execution of the algorithm, we have:
 \begin{equation}
     |I| \leq n/\ell
 \end{equation}

 Observe that $|\cS| \leq t-1$, and each supernode $\rho\in \cS$ contains a skeleton $T_{\rho}$ such that it has at most $t-1$ leaves, and each root-to-leaf path has length at most $(2\ell \log(n)+1)$. Thus:
 \begin{equation}\label{eq:skeleton-size}
     |T_{\rho}|\leq 1 + (t-1)(2\ell \log n+1)~,
 \end{equation}
which means $\rho$ can be broken into  $ O(t\ell\log n)$ stars. By \Cref{lm:ghost-bound}, $|\cX|\leq t(t-1)$, giving:

 \begin{equation*}
    |\cF| \leq |I| + |\cX| + |\textsc{BreakStar}(\cS)|  = O(n/\ell + t^2\ell \log n)
 \end{equation*}
as claimed in \Cref{eq:sep-size}.

\paragraph{Correctness.} Here we show that $\cF$ is a balanced separator of $G$. First, we prove that $\cF$ separates $C$ from other vertices of the graph in every iteration of the algorithm.

\begin{lemma}\label{lm:cF-separate}
Let $(\cS,\cX,I,C)$ be the tuple at the beginning of any iteration of the algorithm. Let $V(\cS,\cX,I)$ be the set of all the vertices of $G$ that are in $I$ or in the supernodes of $\cS\cup \cX$. Then there is no edge in $G$ between $C$ and $V\setminus (C\cup V(\cS,\cX,I))$.  
\end{lemma}
\begin{proof}
    We prove by induction; the base case holds by the construction. Now we show that after an iteration, the updated tuple for the next iteration satisfies the lemma. 

    First, observe that the construction in both Case 1 and Case 2 only moves vertices from $C$ to $V(\cS,\cX,I)$ via creating new supernodes or adding vertices to $I$ in Case 2b.  No vertex in $C$ has been moved outside of $C\cup V(\cS,\cX,I)$ except for when $C$ is updated to be the largest connected component of $C - \kappa$ (Case 1), $C - \rho$ (Subcase 2a), or $C - L_{i}$ (Subcase 2b). However, since the updated $C$ is a connected component, there are no edges in $G$ between the updated $C$ and other discarded connected components. Now consider the \hyperlink{CleaningStep}{cleaning procedure}. Step 1(i) and step 3(i) move vertices out of $C\cup V(\cS,\cX,I)$ by removing either a connector from $\cX$ or a router from $\cS$. In step 1(i) and step 3(i), the condition for the removal is that the supernode has no edge to $C$. In step 1(ii), every vertex of $\kappa \cap N_{G}(C)$ is reassigned and removed from $\kappa$ before we remove $\kappa$ from $\cX$, and hence this updated $\kappa$ is not adjacent to $C$ in $G$. Similarly, in step 3(ii), every vertex of $\rho \cap N_{G}(C)$ is reassigned and removed from $\rho$ before we remove $\rho$ from $\cX$, and hence this updated $\rho$ is not adjacent to $C$ in $G$. This completes the proof.
\end{proof}

\Cref{lm:cF-separate} shows that $\cF$ is a separator. We now show that $\cF$ is balanced. Let $(\cS,\cX,I,C)$ be the input tuple in the last iteration where the size of $C$ remains at least $2n/3$. Let $\check{C}$ be the largest component of $C -L_i$ (Subcase 2b) or $C-\eta$ immediately before the cleaning.  Since the cleaning step does not change $\check{C}$, we have $|\check{C}| \leq 2n/3$. Since $L_i$ is added to $I$ and $\eta$ is added to $\cX\cup\cS$, at this point, every connected component of  $G\setminus V(\cS,\cX,I)$ must have size at most $2n/3$, implying that $\cF$ is balanced. This also completes the proof of \Cref{thm:separator}.

A corollary of our proof:

\begin{corollary}\label{cor:sep-isolated} Let $G$ be a graph with  $n$ vertices that excludes $\dot K_t$ as an induced minor. Let $\ell \geq 1$ be any parameter. Then there exists a balanced separator in $G$ that consists of $n/\ell$ vertices and $O(t^2\ell \log(n))$ stars that can be found in polynomial time. Furthermore, our separator can be made balanced w.r.t. any non-negative weight function $w: V\rightarrow \real_{\geq 0}$ on the vertices, where each connected component has total vertex weight at most $2W/3$ where $W = \sum_{v\in V}w(v)$.
\end{corollary}

\section{A Sample of Applications} \label{sec:applications}

In this section, we present some applications of \Cref{thm:separator-H}.

\subsection{Lazy cop number}\label{subsec:LazyCop}
In the standard version of the cop-and-robber game, any number of cops can move during a single turn, while in the \EMPH{lazy} version, at most one cop can move. Recall that in the standard version, the cop number of a connected graph $G$, denoted by $c(G)$, is the minimum number of cops such that the cop player has a winning strategy. Similarly, in the lazy version, the \EMPH{lazy cop number} of a connected graph $G$, denoted \EMPH{$c_L(G)$}, is the minimum number of cops such that the cop player has a winning strategy. Observe that $c(G)\leq c_L(G)$ for any connected graph $G$.

Compared to the standard version of the cop-and-robber game, the lazy version is much less understood, and its behavior can be substantially different. For example, the \EMPH{Meyniel conjecture} asserts that $c(G)=O(\sqrt{n})$ for every connected $n$-vertex graph~\cite{Baird_2012}.
In contrast, Bal, Bonato, Kinnersley, and Pra{\l}at~\cite{BAL_2015} proved that the hypercube $Q_d$, which has $n=2^d$ vertices, satisfies
$  c_L(Q_d)
    =\tilde{\Omega}(n)$.
Thus, the lazy cop number can be nearly linear in the number of vertices, ruling out an analogue of Meyniel's conjecture for the lazy version. As another example, the argument of Andreae~\cite{Andreae86} mentioned in \Cref{sec:intro} and \Cref{subsec:overview} does not directly apply to the lazy version (as Andreae~\cite{Andreae86} used cops that guard entire  shortest paths, by moving in response to the robber, which is not allowed in the lazy game).

Using a recursive separator argument, Bal, Bonato, Kinnersley, and Pra{\l}at~\cite{Bal_2016} proved that $c_L(G)=O(\sqrt{(g+1)n})$ for every connected graph embeddable on an orientable surface of genus~$g$. For fixed genus, this gives a square-root upper bound, separating these graphs from the above-mentioned hypercubes with nearly linear lazy cop number. The same argument, together with the recent 
$\tilde{O}(h)\cdot \sqrt{n}$-vertex separator theorem of Le \cite{L26}, implies that
$c_L(G)=\tilde{O}(h)\cdot \sqrt{n}$ 
for every connected $H$-minor-free graph $G$, where $h=|V(H)|$. 
However, even for planar graphs, it is not known whether the lazy cop number is bounded by an absolute constant~\cite{Bonato_2020}. Using our \Cref{thm:separator-H}, we obtain the following bound for the much broader class of graphs excluding an $h$-vertex induced minor.

\begin{theorem}\label{thm:lazy-cops} For every graph $H$ with $h$ vertices, every connected $H$-induced-minor-free $n$-vertex graph $G$ satisfies
$c_L(G)=O(h\sqrt{n\log n})$.
Moreover, the corresponding winning strategy can be executed in polynomial time.
\end{theorem}

\begin{proof}[Proof sketch]
We follow the recursive separator argument of Bal, Bonato, Kinnersley, and Pra{\l}at~\cite{Bal_2016}. We use one stationary cop to guard an entire star. The intuition is that if a cop occupies the center of a star and the robber enters its closed neighborhood, this cop captures the robber on the next turn.

Suppose that the robber is confined to a connected induced subgraph $F$ of $G$. By \Cref{thm:separator-H}, $F$ has a balanced separator consisting of $O(h\sqrt{|V(F)|\log n})$ stars. Keeping all previously placed guards stationary, we move fresh cops, one at a time, to the centers of these stars. This is possible since $G$ is connected. During this deployment, the robber may cross the new separator, but it cannot cross a previously guarded separator without being captured.

Once all new guards are in place, the cop player (which controls the cops) lets the robber make its next move. If the robber lies in a guarded star, the cop player captures it. Otherwise, the robber belongs to a connected component of $F$ minus the new separator containing at most $2|V(F)|/3$ vertices. We recurse on this component, keeping all existing guards in place. Thus, the total number of cops required is at most
\[
    \sum_{i\geq 0}
    O\left(h\sqrt{(2/3)^i n\log n}\right)
    =O(h\sqrt{n\log n}).
\]
The robber's component keeps shrinking, so the strategy guarantees that the cop player eventually captures it. Finally, the strategy takes polynomial time: the separators can be constructed in polynomial time, and each new cop can reach its assigned center along a path of length at most $n-1$.
\end{proof}

\subsection{Distance oracles with additive stretch $2$}
\label{sec:metric-oracle}

Given an (unweighted) graph $G = (V,E)$, a \EMPH{distance oracle with additive stretch $k$}  is a compact data structure $\mathcal{D}$ that, given any two vertices $u,v\in V$, returns an estimate $\widehat d(u,v)$ such that:
\begin{equation}
    d_G(u,v) \leq \widehat d(u,v) \leq d_G(u,v) + k
\end{equation}
The problem is to construct a distance oracle that has small \EMPH{space} (in terms of machine words), \EMPH{query time} (the time it takes to output $\widehat d(u,v)$), and \EMPH{small stretch}. Often there are trade-offs between these parameters. Here we focus on the stretch regime $k = 2$.

For string graphs,  de Berg, Jansen, and Lamme~\cite{BJL26} constructed a distance oracle with additive stretch $2$ that has $O(n^{5/3}\log^{2/3}n)$ space and $O(n^{2/3}\log^{2/3}n)$ query time. Using our \Cref{thm:separator-H}, we can construct a substantially better distance oracle with the same stretch. Furthermore, our oracle can be constructed for any induced-minor-free graphs (including string graphs), which is much broader than graphs studied by de Berg, Jansen, and Lamme~\cite{BJL26}.

\begin{theorem}\label{thm:additive-oracle} Let $H$ be a graph with $h$ vertices, and let $G$ be an $H$-induced-minor-free graph with $n$ vertices. We can construct a distance oracle for $G$ with stretch $2$, space $\Tilde{O}(hn^{3/2})$, and query time $\Tilde{O}(h\sqrt{n})$. The oracle can also be constructed in polynomial time. 
\end{theorem}
\begin{proof} We basically follow the idea of de Berg, Jansen, and Lamme~\cite{BJL26}, which is based on~\cite{AdBT25}. The general idea is as follows: \begin{enumerate}
    \item  Compute a balanced star-based separator $\cS$ of $G$. For each star $S\in \cS$, denote by $c_S$ its center vertex; we compute and store the (exact) distance from $c_S$ to every vertex in $G$.
    \item For each connected component $C$ of $G\setminus \bigcup_{S \in \cS}V(S)$, we recursively apply the step above, unless $|V(C)| \leq \log(n)$, in which case we stop and store the exact distances between every pair of vertices in $C$.
\end{enumerate}
The above process induces a recursion tree $\cT$, where each node corresponds to an induced subgraph of $G$ (the root corresponds exactly to $G$), and each leaf is where the recursion stops. Observe that since the size of each subgraph is reduced by at least $1/3$ in each step, $\cT$ has height $O(\log n)$.

To bound the space, we observe that for each vertex $v\in G$, we store the distances from $v$ to the centers of the stars in the balanced star-based separators of subgraphs with exponentially decreasing size $n$, $2n/3$, $(2/3)^2n,\ldots, (2/3)^in,\ldots$. Thus, the total number of stored distances per vertex $v$ is at most:
\begin{equation}
    \sum_{i=0}^{+\infty} O(h\sqrt{(2/3)^in\log n}) = O(h\sqrt{n\log n})
\end{equation}
Thus, the total space is $O(hn^{3/2}\sqrt{\log n})  = \Tilde{O}(hn^{3/2})$. 

To answer a distance query between $u$ and $v$, we first compute: $\widehat d_G(u,v) = \min_{S\in \cS}\left(d_G(u,c_S) + d_G(v,c_S)\right)$ by simply looking up the two distances for each center $c_S$. Then if $u$ and $v$ are in two different components of $G\setminus \bigcup_{S \in \cS}V(S)$, we return $\widehat d_G(u,v)$. Otherwise, let $C$ be the connected component containing both $u$ and $v$;  we recursively compute the estimate $\widehat d_C(u,v)$  and return $\widehat d(u,v) = \min(\widehat d_G(u,v), \widehat d_C(u,v))$.  Observe that the total query time is:
\begin{equation*}
    O(\sum_{i=0}^{+\infty} h\sqrt{(2/3)^i n }\sqrt{\log n}) = O(h\sqrt{n\log n})
\end{equation*}
as claimed. 

The additive stretch is $2$ because the shortest path from $u$ to $v$ might only pass through a leaf $x$ of a star $S$, but since we return $d_G(c_S,u) + d_G(c_S,v)$ as the estimate, by the triangle inequality, the estimate is at most $d_G(u,v)+2$; we refer readers to~\cite{AdBT25} for a more detailed analysis of the stretch.  

The total construction time is polynomial since the separator can be constructed in polynomial time.
\end{proof}

\paragraph{Distance labeling.}
A \EMPH{distance labeling scheme} is a distributed counterpart to a distance oracle. Given an (unweighted) graph $G = (V,E)$, we assign each vertex $v\in V$ a label $L(v)$, such that given only the labels $L(u),L(v)$ of any two vertices, we can return an estimate $\widehat d(u,v)$ satisfying
$d_G(u,v)\leq \widehat d(u,v)\leq d_G(u,v)+k$.
The problem is to construct a distance labeling scheme that has small \EMPH{label size} (in bits), small \EMPH{decoding time}, and small \EMPH{stretch}.
Distance labeling schemes have been extensively studied, both for general graphs and for restricted graph families; see, e.g.,~\cite{GKKPP01,GPPR04,Dragan_2006,GKU16}.
The distance oracle of de Berg, Jansen, and Lamme~\cite{BJL26} converts directly into a distance labeling scheme with additive stretch $2$ that has $O(n^{2/3}\log^{2/3}n)$ label size and $O(n^{2/3}\log^{2/3}n)$ query time.

Similarly, our distance oracle in \Cref{thm:additive-oracle} also provides a distance labeling scheme. Indeed, for each vertex $v$, we store in its label the distances to all separator centers encountered in the recursion, together with their identifiers, and the exact distances stored for $v$ in its leaf, if any. The query algorithm can then be implemented in the same manner as \Cref{thm:additive-oracle}, using only the two labels.
We obtain the following corollary.

\begin{corollary}\label{cor:additive-labeling}
Let $H$ be a graph with $h$ vertices, and let $G$ be an $H$-induced-minor-free graph with $n$ vertices.
We can construct a distance labeling scheme for $G$ with additive stretch $2$, label size
$O(h\sqrt{n}\log^{3/2}n)=\Tilde{O}(h\sqrt{n})$ bits, and decoding time
$O(h\sqrt{n\log n})=\Tilde{O}(h\sqrt{n})$.
The scheme can also be constructed in polynomial time.
\end{corollary}

\subsection{Spanning tree covers of additive stretch $2$}

Given a graph $G = (V,E)$, a \EMPH{spanning tree cover} of $G$ is a collection $\cT$ of spanning trees of $G$. We say that a spanning tree cover  $\cT$ has additive stretch $k$ if for every two vertices $u\not= v\in V$, there exists a tree $T\in \cT$ such that:
\begin{equation*}
     d_T(u,v)\leq d_G(u,v)+k
\end{equation*}

Spanning tree covers have been studied extensively in the literature~\cite{GKR04,AN19,ACEFN20,ES27} (see \cite{ES27} for a recent overview). Notably, for planar graphs, one can construct a spanning tree cover with $O(\sqrt{n})$ trees that preserves distances exactly~\cite{GKR04}. Here we show that our separator in \Cref{thm:separator-H} gives a spanning tree cover with $\Tilde{O}(h\sqrt{n})$ trees and additive stretch $+2$. Then we show an application of our spanning tree cover to construct a path-reporting distance oracle.

\begin{theorem}\label{thm:tree-cover} Let $H$ be a graph with $h$ vertices, and let $G$ be an $H$-induced-minor-free graph with $n$ vertices. We can construct a spanning tree cover for $G$ with $\Tilde{O}(h\sqrt{n})$ trees and additive stretch $2$ in polynomial time. 
\end{theorem}
\begin{proof} We follow the strategy in the proof of \Cref{thm:additive-oracle}:    we construct a star-based separator $\cS$ for $G$ and for each star $S\in \cS$, we construct a BFS tree rooted at $c_S$, the center of $S$, and add the tree to the tree cover. The number of trees added to the tree cover at this step is $\Tilde{O}(h\sqrt{n})$. Then we continue recursively for each connected component $C$ of $G\setminus \bigcup_{S \in \cS}V(S)$, each of size at most $n_1 = 2n/3$. However, we will not add the trees from each connected component separately to the tree cover, since the number of trees could be big, proportional to the number of connected components, which could be as big as $\Omega(n)$. The idea here is to combine these trees together into a (bigger) spanning tree cover of $G$ as these connected components are vertex-disjoint. As a result, the number of spanning trees of $G$ obtained by combining the tree covers of the components is at most  $ \Tilde{O}(h\sqrt{n_1})$.  Thus the total number of trees is bounded by
\begin{equation*}
    \sum_{i\ge0} O(h\sqrt{(2/3)^in\log n}) = O(h\sqrt{n\log n})
\end{equation*}
The analysis of the additive stretch is exactly the same as in the proof of \Cref{thm:additive-oracle}. 
\end{proof}

\paragraph{Application to path reporting distance oracle. } A \EMPH{path-reporting distance oracle} is a data structure that, for every two queried vertices $u,v$, returns, in addition to the distance estimate $\widehat d(u,v)$, a path $P_{uv}$ between $u$ and $v$ in $G$. The oracle has query time $Q(n)$ if it can return $P_{uv}$ in time $Q(n) + O(|P_{uv}|)$.  

Our spanning tree cover in \Cref{thm:tree-cover} implies a path-reporting distance oracle with $\Tilde{O}(hn^{3/2})$ space, additive stretch $2$, and query time $\Tilde{O}(h\sqrt{n})$. Generally, given a spanning tree cover $\cT$ with $t$ trees and additive stretch $k$, we can construct a path-reporting distance oracle with $O(tn)$ space, additive stretch $k$, and query time $O(t)$ as follows. 

For each tree $T\in \cT$, it is well known that we can construct an (exact) distance oracle $\cD_T$ in $O(n)$ space and $O(1)$ query time; see e.g.~\cite{ES27}. We also root each tree at an arbitrary vertex.  The final oracle contains (rooted) $\cT$ and all the tree oracles $\{\cD_T\}_{T\in \cT}$; the total space is $O(tn)$.  To answer a distance query between $u$ and $v$, we  first query the distance $d_T(u,v)$ from each tree $T\in \cT$ using $\{\cD_T\}_{T\in \cT}$, and choose the tree, say $T^*$, with the minimum distance; this step takes $O(t)$ time. Then we find the (unique) path $P_{uv}$ from $u$ to $v$ in $T^*$ in $O(|P_{uv}|)$ time by walking up the tree from $u$ and $v$ to the root simultaneously, and marking visited vertices along the way. As soon as we encounter the first vertex $w$ visited twice, we obtain $P_{uv}$, the path from $u$ to $w$ and then from $w$ to $v$ in $T^*$. Thus, the total time to return the path is $O(t + |P_{uv}|)$ and hence the query time is $O(t)$.  The additive stretch of the oracle is $k$ since $\cT$ has additive stretch $k$.

\subsection{Additive embeddings into low-treewidth graphs (and their applications)}
\label{sec:ai-lowtw-emulator}
Embeddings from planar and minor-free graphs into low-treewidth graphs have
been studied with both additive \cite{FKS19,CFKL20,FL21,FL22}
and multiplicative \cite{CLPP23,CCCLPP25} distortions.
Given such an embedding, one can solve a metric optimization problem in the host graph
using dynamic programming, and then transfer the solution back to the
input graph while controlling the change in distances.
Given an unweighted graph $G=(V,E)$, an \EMPH{embedding with additive
stretch $\beta$} into a weighted graph $G'$ is a function
$f:V\to V(G')$ such that
\[
    d_G(u,v)\leq d_{G'}(f(u),f(v))\leq d_G(u,v)+\beta
    \qquad\text{for every }u,v\in V.
\]
For string graphs of diameter $D$, Chang, Conroy, Tan, and
Zheng~\cite{Chang_2026} constructed an embedding with additive
stretch $\varepsilon D+O(1)$ into a graph of treewidth
$O(\varepsilon^{-4})$.
Using our star-based separators, we obtain additive stretch $2$
and treewidth $O(h\sqrt{n\log n})$ for every
$H$-induced-minor-free graph, where $h=|V(H)|$.
The embedding uses the original vertex set and can be constructed
in polynomial time. The dependence on $n$ is optimal up to a
factor of $\sqrt{\log n}$ in general, already for planar graphs
(see the lower-bound paragraph below).

\begin{theorem}\label{thm:ai-lowtw-emulator}
Let $H$ be a graph with $h$ vertices, and let $G$ be an unweighted $H$-induced-minor-free graph with $n$ vertices.
In polynomial time, we can construct an embedding with additive
distortion $2$ into a weighted graph $G'$ on $V(G)$ of treewidth
$O(h\sqrt{n\log n})$.
\end{theorem}

\begin{proof}
We construct the embedding and a tree decomposition recursively.
For a single vertex, the identity embedding and a single bag suffice.
If $G$ is disconnected, we can embed each connected
component separately.
Hence assume that $G$ is connected and $n\geq2$.
Apply \Cref{thm:separator-H}, and let $S$ be the set of centers
of the separator stars. Thus $|S|=O(h\sqrt{n\log n})$, and
every connected component $G_i$ of $G-N_G[S]$ has at most
$2n/3$ vertices. Recursively construct a host graph $G_i'$ and
a tree decomposition for each $G_i'$.
To construct $G'$, take the disjoint union of the graphs $G_i'$,
add the vertices of $N_G[S]$, and connect every vertex of $S$
to every other vertex of $G$. In particular, vertices in
$N_G[S]\setminus S$ are adjacent only to vertices of $S$.
Give every new edge $uv$ of $G'$ weight $d_G(u,v)$.

We next construct a tree decomposition of $G'$.
Add $S$ to every bag of each recursive decomposition, create a
central bag $S$, and connect one bag from each recursive
decomposition to the central bag. For every
$v\in N_G[S]\setminus S$, attach a leaf bag $S\cup\{v\}$
to the central bag. 
Clearly, we've obtained a proper tree-decomposition for $G'$ of width at most
$|S|+\max\{0,\max_i q_i\}$, where $q_i$ is the respective width of $G_i'$.
As each recursive piece has at most two thirds as many
vertices as its parent, the width is at most
\[
    \sum_{j\geq0}O\!\left(h\sqrt{(2/3)^j n\log n}\right)
    =O\!\left(h\sqrt{n\log n}\right).
\]

It remains to bound the distortion.
Every edge of $G'$ has weight equal to the distance between its
endpoints in $G$. The triangle inequality therefore implies
$d_{G'}(u,v)\geq d_G(u,v)$ for every pair $u,v$.
For the other direction, fix distinct vertices $u,v$ and a
shortest $u$--$v$ path $P$ in $G$.
We distinguish two cases:
\begin{itemize}
    \item If $P$ avoids $N_G[S]$, it is contained in a single component
$G_i$, and hence by the inductive hypothesis, $ d_{G'}(u,v)\leq d_{G_i'}(u,v)
    \leq d_{G_i}(u,v)+2=d_G(u,v)+2$.
    
    \item Otherwise, let $x\in V(P)\cap N_G[S]$ and let $s\in S$ be a center where $x\in N_G[s]$. As both the edges $\{u,s\},\{v,s\}$ belong to $G'$, we have
    $d_{G'}(u,v)
    \leq d_G(u,s)+d_G(s,v)
    \leq d_G(u,x)+d_G(x,v)+2
    =d_G(u,v)+2$.
\end{itemize}
\end{proof}

\paragraph{Lower bound.}
The dependence on $n$ is optimal up to a factor of $\sqrt{\log n}$
in general, already for planar graphs.
The lower-bound construction of Filtser and
Le~\cite{FL22} (Theorem~4 in the full
version) implies that, for every fixed additive error $\beta$,
some unweighted $n$-vertex planar graphs require host treewidth
$\Omega(\sqrt n/\beta)$, even when Steiner vertices are allowed.

\paragraph{Further applications.}
Using \Cref{thm:ai-lowtw-emulator}, we obtain
$2^{\tilde O(h\sqrt n)}$-time additive approximation algorithms
for the following metric optimization problems on 
unweighted $H$-induced-minor-free graphs. These are black-box
reductions to exact algorithms on graphs of bounded treewidth:
we solve the corresponding problem in $G'$ and transfer its
solution to $G$, using the original vertex set and the
additive-two distance guarantee.
\begin{itemize}
    \item \textbf{$\rho$-dominating set.}
    A $\rho$-dominating set contains a vertex within distance $\rho$
    of every vertex of the graph. Using the algorithm of
    Katsikarelis, Lampis, and Paschos~\cite{KLP19}, we find a
    $(\rho+2)$-dominating set whose cardinality is at most the
    minimum cardinality of an optimal $\rho$-dominating set.
    The same algorithm also provides a $+2$ approximation to the \EMPH{$k$-center} problem.

    \item \textbf{$\rho$-independent set.}
    A set is $\rho$-independent if all its pairwise distances are at
    least $\rho$. For $\rho\geq3$, the algorithm of Katsikarelis,
    Lampis, and Paschos~\cite{KLP22} gives a $(\rho-2)$-independent
    set whose weight is at least the maximum weight of a
    $\rho$-independent set.

    \item \textbf{$k$-Median.}
    Each vertex $v$ is a client with nonnegative weight $\mu(v)$,
    and every vertex may be selected as a center. We seek a set
    $C\subseteq V(G)$ of at most $k$ centers minimizing the total
    connection cost $\sum_{v\in V(G)}\mu(v)\,d_G(v,C)$. 
    Thus each client pays its weight times its distance to the
    nearest center. Applying the algorithm of Chakrabarty,
    Conroy, and Sarkar~\cite{CCS25} to $G'$ gives a solution
    whose cost in $G$ is at most $\mathrm{OPT}+2W$, where
    $W=\sum_{v\in V(G)}\mu(v)$ is the total client weight.
\end{itemize}

\subsection*{Acknowledgements}

Research partially completed at the workshop `\href{https://www.matrix-inst.org.au/events/global-structure-and-geometry-of-graphs/}{Global Structure and Geometry of Graphs}' held in April 2026 at MATRIX in Creswick, Australia. We thank all the workshop participants who contributed to the early stage of this work.

\bibliographystyle{alphaurl}
\bibliography{ref,AIbib}

\end{document}